\documentclass[12pt]{article}
\usepackage{cite}
\usepackage[dvips]{graphicx}
\usepackage{amssymb}
\usepackage{rotating}
\usepackage{mathtools}
\usepackage{color,soul}
\usepackage{tikz}

\def \G {\mathcal{G}}
\def \GSM {\G_{SM}}

\def \F {\mathcal{F}}
\def \tminus {\text{-}}

\begin{document}

\vspace*{-1in}
\renewcommand{\thefootnote}{\fnsymbol{footnote}}

\vskip 5pt

\begin{center}
{\Large{\bf 
Surveying the SO(10) Model Landscape:\\ The Left-Right Symmetric Case}}
\vskip 25pt

{
Frank F. Deppisch$^1$\footnote{E-mail: f.deppisch@ucl.ac.uk},
Tom\'as E. Gonzalo$^2$\footnote{E-mail: t.e.gonzalo@fys.uio.no} and
Lukas Graf$^1$\footnote{E-mail: lukas.graf.14@ucl.ac.uk }
}

\vskip 10pt

{\it \small $^1$Department of Physics and Astronomy, University College London, London WC1E 6BT, United
Kingdom}\\
{\it \small $^2$Department of Physics, University of Oslo, N-0316 Oslo, Norway}

\medskip

\begin{abstract}
Grand Unified Theories (GUTs) are a very well motivated extensions of the Standard Model (SM), but the landscape of models and possibilities is overwhelming, and different patterns can lead to rather distinct phenomenologies. In this work we present a way to automatise the model building process, by considering a top to bottom approach that constructs viable and sensible theories from a small and controllable set of inputs at the high scale. By providing a GUT scale symmetry group and the field content, possible symmetry breaking paths are generated and checked for consistency, ensuring anomaly cancellation, SM embedding and gauge coupling unification. We emphasise the usefulness of this approach for the particular case of a non-supersymmetric SO(10) model with an intermediate left-right symmetry and we analyse how low-energy observables such as proton decay and lepton flavour violation might affect the generated model landscape.
\end{abstract}

\end{center}

%\noindent
%{\small PACS numbers:  \\
%\small Keywords: Physics beyond the Standard Model}

\renewcommand{\thefootnote}{\arabic{footnote}}
\setcounter{footnote}{0}

%---------------------------------------------------------
\section{Introduction}
\label{sec:introduction}
%---------------------------------------------------------

Since their first appearance in 1974 \cite{Pati:1974yy, Georgi:1974sy, Georgi:1974yf}, Grand Unified Theories (GUTs) have been one of the most attractive extensions of the Standard Model (SM). The apparent approximate unification of the three gauge couplings in the SM remains an intriguing hint of a unifying origin of gauge interactions around a high energy scale of $10^{16}$~GeV. One such theory is based on the simple Lie group $SO(10)$ \cite{Fritzsch:1974nn, Georgi:1974my} and today remains as one of the preferred candidates for a unified model. There are however a lot of degrees of freedom while building an $SO(10)$ based GUT due to its multiple possible breaking patterns to the SM group and the broad range of representations to choose from. A large number of models have been studied thoroughly and in detail over the years, which has recently resulted in a revivification of the minimal non-supersymmetric $SO(10)$ scenarios~\cite{Galli:1992tf,Bertolini:2010ng,Bertolini:2013vta,Graf:2016znk,Kolesova:2014mfa}. In this work we will take a slightly different approach and construct a rather large set of models based on a specific extended scalar particle content at the $SO(10)$ GUT scale, and we will make a rough survey of cases that represent potentially valid and realistic scenarios as extensions of the SM in terms of several theoretical and phenomenological constraints.

In the present work, we focus on a specific gauge breaking chain,
\begin{align}
	SO(10) \underset{M_{GUT}}{\to}
	SU(3)_C \times SU(2)_L \times SU(2)_R \times U(1)_{B-L} \underset{M_{LR}}{\to}
	\text{SM}
\end{align}
with $M_{GUT}$ denoting the $SO(10)$ GUT breaking scale, with an intermediate Left-Right symmetric gauge group \cite{Pati:1974yy, Mohapatra:1974gc, Senjanovic:1975rk, Duka:1999uc,Maiezza:2016ybz,Chakrabortty:2016wkl} that is subsequently broken to the SM at $M_{LR} < M_{GUT}$. We consider a non-supersymmetric $SO(10)$ realisation. While low energy supersymmetry is a natural companion in GUT considerations~\cite{Anderson:1993fe,Raby:2003in,Raby:2007yv,Blazek:2002ta, Blazek:2001sb,Dimopoulos:1981yj,Dimopoulos:1981zb,Ibanez:1981yh,Sakai:1981gr,Fukuyama:2004ps,Deppisch:2007xu,Deppisch:2014aga}, improving the gauge coupling unification considerably with a minimal particle content and stabilizing the Higgs mass, we want to address the question of what next-to-minimal set of scalar fields can accommodate successful gauge unification while satisfying basic phenomenological constraints.

We start from a fixed particle content at the $SO(10)$ scale with additional Higgs representation but with no exotic fermions beyond the SM leptons and quarks, and the SM sterile neutrino contained in the $\mathbf{16}$-plet. While requiring basic theoretical considerations such as the potential viability to break a given gauge symmetry with a given particle content, we integrate out various representations at the $SO(10)$ and LR-symmetry scale to construct possible models and require gauge unification at the $SO(10)$ scale to determine $M_{LR}$ and $M_{GUT}$. We estimate the predictions for various experimental probes, primarily proton decay and rare lepton flavour violating decays to illustrate how the models can be probed phenomenologically.

Clearly, the construction of realistic models requires more, especially the determination of the scalar potential and a detailed analysis of the symmetry breaking. Given the large number of scenarios considered, we omit this. We also make the simplifying assumption that all states have masses of the associated breaking scale and we neglect the possibility of intermediate particle mass scales. Beyond the gauge structure, we do not consider any other potential symmetries such as discrete or non-compact groups which might be useful to obtain precise gauge coupling unification \cite{Deppisch:2014zta} or to provide a proper flavour structure \cite{King:2014iia}, among other applications.

In our analysis, we do not address a number of fundamental problems \cite{Raby:2007yv} such as fitting of all fermion masses and mechanisms for doublet-triplet splitting. Solutions, maybe partial in general, might be approached through the extended Higgs
sectors enabled in our model setup. Extended Higgs sectors have for example been shown to accommodate fermion masses \cite{Aulakh:2007ir, Aulakh:2006vj, Aulakh:2006hs, Grimus:2006rk}. Alternatively, higher-dimensional operators, suppressed by the Planck-scale can modify the fermion mass patterns \cite{Wiesenfeldt:2005zx}. When such mechanisms are included, also quantitatively, it should be straightforward to translate our results to any such system.

Although we focus only on a few elements in the high-scale scenario, we nevertheless want to elucidate the potential of such an analysis in exploring high-scale scenarios. Refinements and modifications can be implemented if a deeper theoretical context for solving outstanding problems is specified {\it in toto}. In this sense, the results shown may serve as an intermediate step for the analysis of a more comprehensive $SO(10)$ theory. Given the large number of possible BSM scenarios near the Planck scale, experimental projections for a large number of observables such as neutrino masses, proton decay, lepton flavour violating processes, and cosmological observations -- supplemented by careful theoretical considerations -- will be a valuable ingredient for reconstructing the fundamental high-scale theory.

This paper is organized as follows. In Section~\ref{sec:modellandscape} we will describe our basic process of model construction for a general GUT. Section~\ref{sec:leftrightsymmetry} focusses on the specifics of our model choice of $SO(10)$ breaking via an intermediate LR-symmetry scale. Section~\ref{sec:constraints} discusses the basic theoretical considerations we use to constrain the models while Section~\ref{sec:unification} covers the determination of the symmetry breaking scales based on successful gauge unification. In Section~\ref{sec:phenoconstraints} we analyse a non-exhaustive list of experimental constraints. We present our numerical results in Section~\ref{sec:results} and conclude in Section~\ref{sec:conclusions}.

%---------------------------------------------------------
\section{Creating a Model Landscape}
\label{sec:modellandscape}
%--------------------------------------------------------- 

The endeavour of model building begins by specifying the theory at high energies, i.e. providing its symmetries and field content\footnote{A more formal approach of model building requires formulating the Lagrangian of the theory, but for the purpose of this analysis it is sufficient to identify the symmetries and fields.}. Through spontaneous symmetry breaking the scalar potential of the high energy system develops a minimum for a non-zero value of one of the scalar fields, thus supplying that field with a vacuum expectation value and reducing the symmetries of the system. One can then repeat this process, through a series of breaking steps, until the SM symmetries are achieved, tracing a chain of symmetry groups from the high energy to the SM gauge group, $\GSM = SU(3)_C \times SU(2)_L \times U(1)_Y$. At every step of this breaking chain, the field content is obtained by the decomposition of some of the representations from the previous step in the chain, whereas others acquire masses around that particular breaking scale, thereby being integrated out from the spectrum.

Our starting point will be an $SO(10)$-symmetric unified theory at high energy, which can break down in one or more steps to the SM gauge group. Out of all the maximal subgroups of $SO(10)$, only two of them contain $\GSM$ as a subgroup, $SU(5)\times U(1)$ and $SU(4)\times SU(2) \times SU(2)$, hence there are two main branches of symmetry breaking. However, any subgroup of the maximal subgroups can be an intermediate step of the chain, provided the conditions for the symmetry breaking are satisfied. There are in total 15 different ways to realise the $SO(10) \to \GSM$ reduction, ranging from one-step to four-step breaking chains.

Starting with the fields at the highest scale $M_{GUT}$, the scale of $SO(10)$ symmetry breaking, one can obtain the fields at consecutive steps by decomposing the representations of those fields, until the SM scale is reached. Given the elegant feature of $SO(10)$ allowing to accommodate all the SM fermions and a right-handed neutrino of one generation within the $\mathbf{16}_F$ multiplet, we will make the simplifying assumption that no additional fermions are present. Instead, we assume that all other $SO(10)$ representations present in the theory are scalars. 

According to the Extended Survival Hypothesis (ESH)~\cite{Georgi:1979md, delAguila:1980at}, the Higgs scalars acquire a mass compatible with the pattern of symmetry breaking. This means that at every scale the only surviving scalars are those required to satisfy the remaining symmetry breaking steps, whereas the rest of scalars will be integrated out at the GUT scale or at one of the intermediate scales. This is clearly a minimal choice, as in general these scalar fields are allowed to live at any scale, with masses that will be obtained dynamically from the configuration of the Lagrangian and the couplings involved.

Nevertheless, at this stage of model building we do not know the configuration of the Lagrangian or its couplings. Though this could mean that we have no control over the particle masses, for in general they may not be associated with any scale, here we make the simplifying assumption that fields acquire masses of the order of their associated breaking scale. Thus we will assume, a priori, that all fields have the potential to survive or be integrated out at any of these symmetry breaking scales\cite{Czakon:1999ha,Chankowski:2006hs}, as long as the present scalar fields can in principle achieve the desired breaking pattern. This allows for a very large set of models, particularly when high-dimensional $SO(10)$ representations are considered. Neglecting any other constraints and theoretical requirements, there are simply $2^n$ possible combinations of fields out of the $n$ fields obtained from the decomposition of $SO(10)$ representations. In order to make the analysis more manageable, and inspired by the ESH, we will only keep a small set of fields at every scale. Hence, whenever there is a large number of representations at a certain scale, we will restrict to having only up to $k = 5$ representations out of the whole set at that scale. The number of possible cases is then reduced significantly, and is given by
\begin{equation}
 N = \sum_{k=0}^5
 \begin{pmatrix}
  n \\ k
 \end{pmatrix} \approx 3.8\times 10^5, \quad \text{for } n=35.
\end{equation}
with $n$ denoting the original number of representations at that given scale.

Among the many combinations of representations available, duplicates may appear. Often identical or conjugate representations in the spectrum can originate from more than one $SO(10)$ representation, which results in multiple ways to generate the same field content. Although from a low energy viewpoint they are indistinguishable and have the same RGEs, their different origin means that their Lagrangian terms might be different and thus they are not, from the high energy perspective, the same model. Hence, when generating the models, this degeneracy of the representation content is kept, because they represent different models and it might affect any current or future Lagrangian-level analysis. In addition, gauge singlets have no effect on the 1-loop RGEs or any of the low energy phenomena considered here, yet they are also kept in the models, for the same reasons.

Therefore, the algorithm for generating models, given a set of representations at the $SO(10)$ scale and a symmetry breaking chain, consists of decomposing the fields into subsequent groups in the chain and, after applying the constraints (see Section \ref{sec:constraints} below), obtaining all possible combinations, as discussed above\footnote{For more details about the algorithm of model generation, see~\cite{GonzaloVelasco:2015hki}.}. Repeating this process for all scales, the outcome will be a landscape of models, where each model will be defined by the sets of representations at the different scales. For a chain $\G \to \F_1 \to \dots \to \F_{m-1} \to \GSM$, with $m$ being the number of breaking steps, the set of models will be a list of the type
\begin{equation}
 \{\mathcal{M}\} = \left\{
 \begin{array}{ll}
  \text{Chain: } \{\G ~\to \dots \to \GSM\}, & \text{Reps: } \{\mathcal{R}^{(0)}_i\}, \\
  \text{Chain: } \{\F_1 \to \dots \to \GSM\}, & \text{Reps: } \{\mathcal{R}^{(1)}_i\}, \\
  \quad \dots & \quad \dots\\
  \text{Chain: } \{\GSM\}, & \text{Reps: } \{\mathcal{R}^{(m)}_i\} \\
 \end{array}
 \right\},
 \label{model}
\end{equation}
where $\{\mathcal{R}_i^{(0)}\}$ are the representations at the $SO(10)$ scale and $\{\mathcal{R}^{(j)}_i\}$ a combination of their decompositions at the $j$th step.

\section{$SO(10)$ with intermediate Left-Right Symmetry}
\label{sec:leftrightsymmetry}
%---------------------------------------------------------

As mentioned, there are many possible breaking chains from $SO(10)$, of which the most interesting are those with intermediate product groups. The addition of an intermediate scale generally simplifies the issue of gauge coupling unification, even more so in multiple step breaking chains. As a first approach, we will take the two-step breaking from $SO(10)$ with the left-right symmetry group $SU(3)_C \times SU(2)_L \times SU(2)_R \times U(1)_{B-L}$ at an intermediate scale, because some of its minimal realisations have been analysed extensively in the literature, see e.g.~\cite{Mohapatra:1974gc, Mohapatra:1974hk, Senjanovic:1975rk, Senjanovic:1978ev, Deshpande:1990ip, Aulakh:1997ba, Aulakh:1997fq, Duka:1999uc, Zhang:2007da, Brahmachari:1991np, Brahmachari:2003wv, Siringo:2012bc, Lindner:1996tf, Arbelaez:2013nga, Mohapatra:1986bd, Das:2012ii, Dev:2013oxa, Duerr:2013opa, Deppisch:2015qwa, Deppisch:2014qpa, Deppisch:2014zta, Dekens:2015csa}. Other two-step breaking chains, such as those that go through $SU(5)$, may lead to similar situations with added difficulties such as three-way gauge coupling unification and rapid proton decay. Therefore, we defer the study of these to future work and focus here on models with an intermediate LR scale.

As mentioned before, the main ingredients for the start of the model building process are the gauge group, the breaking chain and the set of representations. We have already chosen our GUT gauge group to be $SO(10)$, and the gauge symmetry breaking chain of this scenario reads
\begin{align}
SO(10) \to  SU(3)_C\times SU(2)_L\times SU(2)_R\times U(1)_{B-L} \to  SU(3)_C\times SU(2)_L\times U(1)_{Y}.
\end{align}

Lastly, the set of $SO(10)$ representations needs to be specified. In general most sets will not produce successful models because some of the symmetry breaking conditions require very specific representations. We thus choose quite a large set of initial representations, partially inspired by previous work on this type of symmetry breaking~\cite{Arbelaez:2013nga}. The chosen set is\footnote{Many of the equations and tables below include $SO(10)$ representations not in this list. These are not included in the analysis, but are kept for reference}.
\begin{equation}
\{3\times \mathbf{16}_F, \mathbf{10}, \mathbf{45}, \mathbf{126}, \mathbf{\overline{126}}\},
\end{equation}
where there are three generations of $\mathbf{16}_F$, which unify the SM fermions; a scalar $\mathbf{10}$ which will contain part of the SM Higgs boson; a $\mathbf{45}$ of adjoint scalars, required for the first step of symmetry breaking at the $SO(10)$ scale; and two 126-dimensional representations that will contain fields responsible for further symmetry breaking steps and a field that can contribute to the SM Higgs boson. The right-handed neutrinos in the $SO(10)$ 16-plets will be part of $SU(2)_R$ doublets in the LR symmetric phase and thereby contribute to the RGE running. We will assume that they acquire heavy Majorana masses of the order of the LR symmetry breaking scale to potentially generate light left-handed neutrino masses of order 0.1~eV in a seesaw mechanism. 

Given the large representations used, their decomposition in the intermediate group $SU(3)_C\times SU(2)_L\times SU(2)_R\times U(1)_{B-L}$ contains many terms,
\begin{align}
\notag \mathbf{10} &\to \{\mathbf{3},\mathbf{1},\mathbf{1},~\tfrac{1}{2}\} \oplus \{\mathbf{\bar{3}}, \mathbf{1}, \mathbf{1},\tminus\tfrac{1}{2}\} \oplus \{\mathbf{1},\mathbf{2},\mathbf{2},0\},\\
\notag \mathbf{45} &\to \{\mathbf{3},\mathbf{2},\mathbf{2},~\tfrac{1}{2}\} \oplus \{\mathbf{\bar{3}},\mathbf{2},\mathbf{2},\tminus\tfrac{1}{2}\} \oplus \{\mathbf{8},\mathbf{1},\mathbf{1},~0\} \oplus \{\mathbf{\bar{3}},\mathbf{1},\mathbf{1},~1\} \oplus \{\mathbf{1},\mathbf{3},\mathbf{1},0\} \\
\notag &~\oplus \{\mathbf{3},\mathbf{1},\mathbf{1},\tminus1\} \oplus \{\mathbf{1},\mathbf{1},\mathbf{3},~0\} \oplus \{\mathbf{1},\mathbf{1},\mathbf{1},~0\},\\
\notag \mathbf{126} &\to \{\mathbf{8},\mathbf{2},\mathbf{2},~0\} \oplus \{\mathbf{6},\mathbf{3},\mathbf{1},\tminus\tfrac{1}{2}\} \oplus \{\mathbf{\bar{6}},\mathbf{1},\mathbf{3},~\tfrac{1}{2}\} \oplus \{\mathbf{\bar{3}},\mathbf{2},\mathbf{2},~1\} \oplus \{\mathbf{3},\mathbf{2},\mathbf{2},\tminus1\} \\
\notag &~\oplus \{\mathbf{3},\mathbf{3},\mathbf{1},~\tfrac{1}{2}\} \oplus \{\mathbf{\bar{3}},\mathbf{1},\mathbf{3},\tminus\tfrac{1}{2}\} \oplus \{\mathbf{1},\mathbf{2},\mathbf{2},~0\} \oplus \{\mathbf{3},\mathbf{1},\mathbf{1},~\tfrac{1}{2}\} \oplus \{\mathbf{\bar{3}},\mathbf{1},\mathbf{1},\tminus\tfrac{1}{2}\} \\
\notag &~\oplus \{\mathbf{1},\mathbf{3},\mathbf{1},~\tfrac{3}{2}\} \oplus \{\mathbf{1},\mathbf{1},\mathbf{3},\tminus\tfrac{3}{2}\}, \\
\notag \mathbf{\overline{126}} &\to \{\mathbf{8},\mathbf{2},\mathbf{2},~0\} \oplus \{\mathbf{\bar{6}},\mathbf{3},\mathbf{1},~\tfrac{1}{2}\} \oplus \{\mathbf{6},\mathbf{1},\mathbf{3},\tminus\tfrac{1}{2}\} \oplus \{\mathbf{3},\mathbf{2},\mathbf{2},\tminus1\} \oplus \{\mathbf{\bar{3}},\mathbf{2},\mathbf{2},~1\} \\
\notag &~\oplus \{\mathbf{\bar{3}},\mathbf{3},\mathbf{1},\tminus\tfrac{1}{2}\} \oplus \{\mathbf{3},\mathbf{1},\mathbf{3},~\tfrac{1}{2}\} \oplus \{\mathbf{1},\mathbf{2},\mathbf{2},~0\} \oplus \{\mathbf{\bar{3}},\mathbf{1},\mathbf{1},\tminus\tfrac{1}{2}\} \oplus \{\mathbf{3},\mathbf{1},\mathbf{1},~\tfrac{1}{2}\} \\
&~\oplus \{\mathbf{1},\mathbf{3},\mathbf{1},\tminus\tfrac{3}{2}\} \oplus \{\mathbf{1},\mathbf{1},\mathbf{3},~\tfrac{3}{2}\}.
\end{align}

A priori, there are 35 scalar representations, so there will be $N = 2^{35} \sim 10^{10}$ possible combinations. As was mentioned before, in order to be able to perform a reasonable quantitative analysis, we will restrict the field content to up to 5 representations above the left-right (LR) scale. The number of combinations of representations is now close to $4 \times 10^5$, a more manageable amount, of which only about $2.5 \times 10^5$ models will satisfy the theoretical constraints laid out in Section \ref{sec:constraints}.

\section{Theoretical Constraints} 
\label{sec:constraints}
%---------------------------------------------------------
Despite the large number of models obtained via the process described above, not all of them will be valid candidates for a GUT. Each of the models, i.e. each of the combinations of fields, must satisfy a set of constraints at every step of the breaking chain, in order to be considered as a successful model. We would like to stress again that we only include a basic set of constraints based on the group breaking structure and the set of representations.

\textbf{Chirality} The gauge group of the theory, at every scale, must allow for its representations to respect the chiral structure of the SM, i.e. that left and right-handed fields transform under conjugate representations of the group. For simple groups this means that the group must allow complex representations, satisfied by unitary groups $SU(n)$, orthogonal groups of the type $SO(2n)$ with odd $n$, and the exceptional algebra $E_6$. Since the $SO(10)$ group is precisely one of the allowed cases for orthogonal groups, it satisfies the condition as long as the SM fermions are embedded in the 16-dimensional representation. The chirality condition is thus satisfied automatically for all steps of the breaking chain, for each of the breaking patterns, because they always involve unitary and semi-simple sub-algebras.

\textbf{Anomalies} Several anomalies can arise in a gauge theory, namely the gauge~\cite{Adler:1969gk,Bell:1969ts}, gravitational~\cite{AlvarezGaume:1983ig} and Witten~\cite{Witten:1982fp} anomalies. For the purpose of our work, the only relevant of these is the gauge or Adler-Bell-Jackiw anomaly. Gauge anomalies occur in theories with massive vector bosons, where triangle diagrams involving fermionic loops do not cancel. In such cases, the symmetry is broken at the quantum level and the theory becomes non-renormalizable~\cite{Gross:1972pv}. The contribution of these diagrams to the anomaly is proportional to $\mathcal{A}_{abc}^{i} = \text{Tr}(\{T_a^i,T_b^i\}T_c^i) $\cite{Georgi:1972bb, Georgi:1999wka}, where $T_a^i$ are the generators of the group or groups associated with the gauge bosons on the external legs, written in the representation of the fermion $f_i$ running inside the loop. 

Most simple Lie algebras are automatically free of this type of anomaly and they are known as safe algebras\cite{Georgi:1972bb}, with the notable exception\footnote{The orthogonal algebra $SO(6)$ is not safe either, since it is isomorphic to $SU(4)$, which is unitary and thus not safe.} of unitary algebras $SU(n)$ for $n\geq3$ and the exceptional algebra $E_6$. In those cases, one must compute the contribution to the anomaly from all the fermions in the theory and require that their sum cancels, $\sum_i \mathcal{A}_{abc}^i = 0$. For non-semi-simple algebras, the gauge bosons in the external legs of the triangle diagram could belong to different factors of the product group. However, the tracelessness and orthogonality properties of the generators $T_a$ ensure that any diagram with different non-Abelian gauge bosons on the external legs cancel automatically~\cite{Peskin:1995ev}.

By construction, all models created in this analysis are anomaly free. This is because there are no exotic fermions in non-trivial representations of the algebra, and the fermionic matter sector is embedded in the anomaly-free $\mathbf{16}$ representation of $SO(10)$.

\textbf{Symmetry Breaking} Since there are one or more symmetry breaking steps throughout the chain, one needs to make sure that these can be realised by the scalar representations present in the theory. Thus, for every step, we will require at least one field in the theory that can break the symmetry to the next step of the chain. This means that the set of representations of a step must contain a non-singlet representation that is a singlet under the chosen subgroup (the group in the next step of the chain). Nevertheless, the existence of such a representation is not enough to trigger the breaking of the symmetry; one also needs to make sure that there is a transition between the symmetry-preserving and symmetry-breaking vacua. However, this would require knowledge of the scalar potential of the theory and the parameters within, which falls out of the scope of this analysis. Therefore, we will consider as a necessary and sufficient condition for symmetry breaking that a representation capable of doing so is present.

\textbf{Standard Model} The last step of the breaking chain in any realistic GUT is the SM gauge group, $SU(3)_C \times SU(2)_L \times U(1)_Y$, so one needs to ensure that at least the SM matter content is reproduced here, including the precise hypercharge assignments (modulo an overall normalisation factor). This condition requires the presence of all three generations of SM fermions in the last step of the chain, and also the existence of at least a Higgs doublet, so as to satisfy electroweak symmetry breaking. In addition, we will potentially allow the presence of extra scalar fields, either singlets or charged under the weak and hypercharge groups (but with no colour charge) that may affect SM physics, but which in general are not immediately ruled out.

\textbf{Fermion mixing} Lastly, one could attempt to reproduce the SM values for fermionic masses and mixing angles, i.e. the CKM matrix~\cite{Joshipura:2011nn,Babu:2016bmy,Meloni:2016rnt}. Though calculating the specific values for the parameters lies beyond the scope of this work, it is possible to set up certain constraints that, at the very least, guarantee enough degrees of freedom to fit the parameters. At the $SO(10)$ level, considering only renormalisable terms, the Yukawa couplings of fermions are given by
\begin{equation}
 \mathcal{L}_{Y} = \mathbf{\bar{16}}_F \left( \mathbf{Y}_{10}  \mathbf{10} + \mathbf{Y}_{120}  \mathbf{120} + \mathbf{Y}_{126}  \overline{\mathbf{126}} \right) \mathbf{16}_F,
\end{equation}
with $\mathbf{Y}_X$ matrices of Yukawa couplings. Thus, if $v_{u,d}$, $\omega^{\alpha,\beta}_{u,d}$ and $\sigma_{u,d}$ are the up and down-type vacuum expection values (VEVs) of $\mathbf{10}$, $\mathbf{120}$\footnote{The two possible VEVs for the $\mathbf{120}$ representation are labelled $\alpha$ and $\beta$, along the direction of one of the two SM doublets it contains.} and $\overline{\mathbf{126}}$, respectively, the masses of the fermions can be written as \cite{DiLuzio:2011my}
\begin{align}
 \notag \mathbf{M}_u &= \mathbf{Y}_{10} ~v_u + \,~\mathbf{Y}_{126} ~\sigma_u + \mathbf{Y}_{120}~(\omega_u^\alpha + \,~~\omega_u^\beta), \\
 \notag \mathbf{M}_d &= \mathbf{Y}_{10} ~v_d + \,~\mathbf{Y}_{126} ~\sigma_d + \mathbf{Y}_{120}~(\omega_d^\alpha + \,~~\omega_d^\beta), \\
 \notag \mathbf{M}_e &= \mathbf{Y}_{10} ~v_d - 3 \mathbf{Y}_{126}~ \sigma_d + \mathbf{Y}_{120}~(\omega_d^\alpha -3 ~\omega_d^\beta), \\
 \mathbf{M}_\nu &= \mathbf{Y}_{10} ~v_u - 3 \mathbf{Y}_{126} ~\sigma_u + \mathbf{Y}_{120}~(\omega_u^\alpha - 3 ~\omega_u^\beta),
 \label{fermionmasses}
\end{align}
where $\mathbf{M}_i$ is the Dirac mass matrix for the fermion species $i$. The combination of fields $\mathbf{10} + \mathbf{\overline{126}}$ can provide a reasonable fit of the fermion masses to their measured values, which for our test case, as described in Section \ref{sec:leftrightsymmetry}, is satisfied by construction. At other steps of the breaking chain it may be necessary to impose other constraints, specific to the particular cases. For the LR symmetric model, our case of interest, it can be proven \cite{Arbelaez:2013nga} that the minimal requirement is two bidoublets $\{\mathbf{1},\mathbf{2},\mathbf{2},0\}$ and a right-handed triplet $\{\mathbf{1},\mathbf{1},\mathbf{3},0\}$.

%---------------------------------------------------------
\section{Gauge Coupling Unification}
\label{sec:unification}
%---------------------------------------------------------

Once we have obtained the set of valid models and applied the constraints above, the next step is to ensure that the breaking chain is consistent with the unification of the gauge couplings. Their running is described by corresponding RGEs, which for each model depend on the representations at a given scale. 

The set of RGEs together with the initial condition imposed by SM couplings at the electroweak scale and by the unification condition at $M_{GUT}$ form a stringent constraint on any GUT symmetry breaking scenario. The values of the gauge couplings at the SM scale are~\cite{Olive:2014kda}
\begin{align}
 \notag g_1(M_Z) &= 0.46235 \pm 0.00010,\\
 \notag g_2(M_Z) &= 0.65295 \pm 0.00012,\\
 g_3(M_Z)        &= 1.220   \pm 0.003,
 \label{SMcouplings}
\end{align}
where $1$, $2$ and $3$ refer to the $U(1)$, $SU(2)$ and $SU(3)$ groups, respectively.

Solving the RGEs is in general a difficult endeavour because they usually depend on the other parameters in the theory and form a system of strongly coupled differential equations. We will restrict our analysis to the one-loop level for which the gauge coupling RGEs are uncoupled and can be easily solved analytically. The one-loop RGE for the gauge coupling $g$ of a group $\G$ is
\begin{equation}
\mu\dfrac{\mathrm{d}g}{\mathrm{d}\mu} =  \frac{b}{16\pi^2}g^3,
\label{gaugerges}
\end{equation}
where $\mu$ is an energy scale and the slope $b$ is calculated as~\cite{Martin:1993zk}
\begin{equation}
b = \frac{2}{3}\sum_{\text{Fermions}} S(\mathcal{R}_f) d_{\bot}(\mathcal{R}_f) + \frac{1}{3} \sum_{\text{Scalars}} S(\mathcal{R}_s) d_\bot(\mathcal{R}_s) - \frac{11}{3} C_2(\G).
\label{smslopes}
\end{equation}
Here, $C_2(\G)$ is the Casimir operator of the group $\G$, $S(\mathcal{R}_{s,f})$ is the Dynkin index of the scalar $\mathcal{R}_s$ or fermionic $\mathcal{R}_f$ representation under the group $\G$ and $d_\bot(\mathcal{R}_{s,f})$ is the number of degrees of freedom of the representation $\mathcal{R}_{s,f}$ under the groups orthogonal to $\G$. For Abelian groups, such as the hypercharge factor in the SM, the Casimir vanishes, $C_2(U(1)) = 0$, and the Dynkin index of a representation with $U(1)$ charge $Q$ is given by $S(Q) = Q^2$.

When a group has more than one Abelian factor, kinetic mixing can lead to coupling of the corresponding RGEs~\cite{Fonseca:2011vn}. This contribution, however, is usually quite small, of the order of the two-loop correction of the RGEs~\cite{Fonseca:2011vn}. In any case, this possibility does not occur in our chosen breaking chain, as will be seen below.

\begin{table}[t!]
 \centering
 \begin{tabular}{c|cccccc}
  \hline
  Field & $\mathcal{R}$ & Spin & $n_f$ & $SU(3)_C$ & $SU(2)_L$ & $U(1)_Y$ \\
  \hline
  $Q$ & $\{\mathbf{3},\mathbf{2},~\tfrac{1}{6}\}$ & $\frac{1}{2}$ & 3 & $\tfrac{1}{2} \cdot 2$ & $3 \cdot \tfrac{1}{2}$ & $3 \cdot 2 \cdot \tfrac{1}{36}$ \\
  $u^c$ & $\{\mathbf{\bar{3}},\mathbf{1},\tminus\tfrac{2}{3}\}$ & $\tfrac{1}{2}$ & 3 & $ \tfrac{1}{2}$ & 0 & $3 \cdot \tfrac{4}{9}$ \\
  $d^c$ & $\{\mathbf{\bar{3}}, \mathbf{1},~\tfrac{1}{3}\}$ & $\tfrac{1}{2}$ & 3 & $ \tfrac{1}{2}$ & 0 & $3 \cdot \tfrac{1}{9}$ \\
  $L$ & $\{\mathbf{1},\mathbf{2},\tminus\tfrac{1}{2}\}$ & $\tfrac{1}{2}$ & 3 & 0 & $\tfrac{1}{2}$ & $2 \cdot\tfrac{1}{4}$ \\
  $e^c$ & $\{\mathbf{1},\mathbf{1},~1\}$ & $\tfrac{1}{2}$ & 3 & 0 & 0 & $1$ \\
  $H$ & $\{\mathbf{1},\mathbf{2},\tminus\tfrac{1}{2}\}$ & 0 & 1 & 0 & $\tfrac{1}{2}$ & $2\cdot\tfrac{1}{4}$\\
  \hline
 \end{tabular}
 \caption[Standard Model particle content and associated properties]{\emph{Standard Model particle content and associated properties: spin, number of families $n_f$ and Dynkin index (times the number of gauge degrees of freedom) under the groups $SU(3)_C$, $SU(2)_L$ and $U(1)_Y$.}}
 \label{tab:smcontent}
\end{table}
As an example, for the SM particle content, Table~\ref{tab:smcontent},\footnote{Throughout this article we have used the Weyl representation for fermionic fields. Fields of the same helicity are contracted with the Levi-Civita symbol $\epsilon$, and fields of opposite helicity with the extended Pauli matrices $\sigma_\mu$ and $\bar\sigma_\mu$. In addition the antisymmetric symbol $\sigma_{\mu\nu} = \tfrac{i}{4}(\sigma_\mu\bar\sigma_\nu - \sigma_\nu\bar\sigma_\mu)$ is used for tensor contractions.} which shows the Dynkin indices of the representations, and with Casimirs $C_2(SU(3)) = 3$, $C_2(SU(2)) = 2$ and $C_2(U(1)) = 0$, one obtains the slopes\footnote{These values of the slopes are calculated including the contribution from the top quark, which we will take to be approximately at the electroweak scale $M_Z$.}
\begin{equation}
 \{b_1,b_2,b_3\} = \left\{\tfrac{41}{10}, -\tfrac{19}{6}, -7\right\},
\end{equation}
for the three SM gauge groups.

The RGEs in Eq.~\eqref{gaugerges} can be conveniently rewritten in terms of the parameter $\alpha^{-1} = (g^2/4\pi)^{-1}$ as
\begin{align}
\mu\frac{\mathrm{d}\alpha^{-1}}{\mathrm{d}\mu}=-\frac{b}{2\pi}.
\label{alpharges}
\end{align}
Changing the variable to $t= 1/2\pi \log (\mu/M_Z)$ and given the boundary condition $\alpha^{-1}(t_0)$ at scale $t_0$, it can be solved analytically as
\begin{align}
 \alpha^{-1}\left(t\right)-\alpha^{-1}\left(t_{0}\right)=-b\left(t-t_{0}\right).
 \label{alphasolution}
\end{align}
For a breaking chain from $SO(10)$ to the SM with $m$ steps, there are $m-1$ intermediate scales $\mu_i$, with $t_i = 1/2\pi \log (\mu_i/M_Z)$. Starting with the unification of gauge couplings at the scale $t_m = t_{GUT} \leftrightarrow \mu_m = M_{GUT}$, the RGEs can be solved at the following scale $\mu_{m-1}$. The new boundary conditions $\alpha(t_{m-1})$ are used to solve for subsequent scales, iterating until the SM scale, $t_0 = 0 \leftrightarrow \mu_0 = M_Z$.

In such a scenario, there are $m+1$ free parameters: the $m-1$ intermediate scales, the GUT scale $M_{GUT}$ and the coupling at the unification scale $\alpha_{GUT}$. On the other hand, the running couplings must match their values at the SM scale, shown in Eq.~\eqref{SMcouplings}, which leaves at least $m-2$ degrees of freedom for any GUT scenario. In our case of two-step breaking, there are no free parameters and the scales and $\alpha_{GUT}$ are all uniquely determined. If further constraints are applied, e.g. if the right-handed current in left-right symmetric models would be observed and thus the associated $SU(2)_R$ gauge coupling was measured, there would be fewer degrees of freedom.

Since Eq.~\eqref{alphasolution} is linear, one can write equations for the SM couplings $\alpha^{-1}_i$, with $i=1,2,3$, that implement the constraint of unification at $\alpha^{-1}_{GUT}$ as
\begin{align}
\alpha^{-1}_{i} = \alpha^{-1}_{GUT} + \sum_{j=1}^{m}b^{i}_{j}\Delta t_{j},
\end{align}
where we have defined the splitting between two consecutive scales as $\Delta t_{j}=t_{j}-t_{j-1}$ with $j=1,\dots,m$, and $b^{i}_{j}$ are the slopes corresponding to particular segments $\Delta t_{j}$ of the path connecting $\alpha^{-1}_{GUT}$ with $\alpha^{-1}_{i}$. One can summarize these three conditions in matrix form as
\begin{align}
\left( \begin{array}{cc}
  \alpha^{-1}_3 \\
  \alpha^{-1}_2 \\
  \alpha^{-1}_1
\end{array} \right)
=
\begin{pmatrix}
  1 & b^{3}_{1} & b^{3}_{2} & \cdots & b^{3}_{m} \\
  1 & b^{2}_{1} & b^{2}_{2} & \cdots &  b^{2}_{m} \\
  1 & b^{1}_{1} & b^{1}_{2} & \cdots &  b^{1}_{m}
\end{pmatrix} \cdot
\begin{pmatrix}
  \alpha_{GUT}^{-1} \\
  \Delta t_{1} \\
  \Delta t_{2} \\
  \vdots \\
  \Delta t_{m}
\end{pmatrix} \equiv B_0 \cdot \Delta t.
\label{matrixeq}
\end{align}

\subsection{Abelian Breaking}
\label{sec:mixing}
%---------------------------------------------------------
In a number of scenarios, namely those where there is a rank-reducing breaking and the subgroup contains an Abelian factor, the generator of the remaining $U(1)$ factor is a linear combination of the diagonal generators of the supergroup. For the simple case $U(1)_A \times U(1)_B \to U(1)_C$, the charges of a field $\phi^j$ under $U(1)_C$ and its gauge coupling can be calculated as
\begin{equation}
   g_C~ Q_C^j = g_A g_B \frac{Q_A^j Q_B^v - Q_B^j Q_A^v}{\sqrt{g_A^2 (Q_A^v)^2 + g_B^2 (Q_B^v)^2}},
\end{equation}
where $g_A$ and $g_B$ are the couplings of $U(1)_A$ and $U(1)_B$, respectively, $Q_A^j$ and $Q_B^j$ are the charges of the field $\phi^j$ and $Q_A^v$ and $Q_B^v$ are the charges of the breaking Higgs. If any or both of the supergroups are not Abelian, then the charges correspond to the eigenvalues of the diagonal generators that survive the breaking.

Though they are not defined independently, we need to use both $g_C$ and $Q_C^j$ separately, the former when solving the RGEs to obtain limits on the scales and the latter to obtain the slopes of the RGEs. We will then choose to define
\begin{equation}
  g_C = g_A g_B \dfrac{\sqrt{(Q_A^v)^2 + (Q_B^v)^2}}{\sqrt{g_A^2 (Q_A^v)^2 + g_B^2 (Q_B^v)^2}} = \dfrac{g_A g_B}{\sqrt{r_A^2 g_A^2 + r_B^2 g_B^2}},
\end{equation}
and
\begin{equation}
 Q_C^j = \dfrac{Q_A^j Q_B^v - Q_B^j Q_A^v}{\sqrt{(Q_A^v)^2 + (Q_B^v)^2}} = r_B Q_A^j - r_A Q_B^j,
\end{equation}
with $r_{A,B} = Q_{A,B}^v/ \sqrt{(Q_A^v)^2 + (Q_B^v)^2}$ such that $r_A^2 + r_B^2 =1$.

In $SO(10)$ unified models this $2\to 1$ Abelian breaking is the only type that will appear, hence the simple analysis above is sufficient. 

At the scale $t_{mix}$ the boundary condition for the gauge coupling in the broken phase $\alpha_C^{-1}$ is given by
\begin{equation} 
 \alpha_C^{-1}(t_{mix}) = r_A^2 \alpha_A^{-1} (t_{mix}) + r_B^2 \alpha_B^{-1}(t_{mix}),
\end{equation}
which allows us to write $\alpha_C^{-1}$ at the EW scale, assuming no dynamical mixing between $U(1)_A$ and $U(1)_B$, as
\begin{align}
\alpha^{-1}_1 = \alpha_C^{-1}(t_0) = \alpha^{-1}_{GUT} + r^2_{A} \sum_{j={mix+1}}^{m}b^{1A}_{j}\Delta t_{j} + r^2_{B} \sum_{j={mix+1}}^{m}b^{1B}_{j}\Delta t_{j} + \sum_{j=1}^{{mix}}b^C_{j}\Delta t_{j},
\label{mixingeq}
\end{align}
where $b^{1A}_j$ and $b^{1B}_j$ correspond to the slopes of the gauge couplings above $t_{mix}$ and $b^C_j = b^1_j$ is the slope of the remaining coupling below $t_{mix}$.

In terms of the matrices in Eq.~\eqref{matrixeq}, one would need three independent matrices of slopes, $B_A$, $B_B$ and $B_C$. The first two have zeroes in every $b^a_j$ entry for $j = 1,\dots, mix$ and the slopes $b^{1A}_j$ and $b^{1B}_j$ for $j = mix+1,\dots, m$. Conversely, the matrix $B_C$ has zero entries on the right side of the mixing scale, $j>mix$ and $b^1_j$ on the left side, $j < mix$. Therefore, the matrix equation takes the form
\begin{equation}
 \begin{pmatrix}
 \alpha^{-1}_3 \\
 \alpha^{-1}_2 \\
 \alpha^{-1}_1
 \end{pmatrix} = (r_A^2 B_A + r_B^2 B_B + B_C) \cdot \Delta t.
 \label{matrixeqmixing}
\end{equation}

\subsection{Solving the RGEs}
%---------------------------------------------------------
The matrix system in Eq.~\eqref{matrixeq} is $\alpha^{-1} = B \cdot \Delta t$, where $\Delta t$ includes the GUT and intermediate scales and the matrix of slopes $B$ is calculated, in case of Abelian mixing, using the structure given in Eq.~\eqref{matrixeqmixing}.

This system of linear equations is solvable for $\Delta t$ when the number of scales $m$, is $m = 2$, which gives unique solutions for the intermediate and the unification scales and $\alpha_{GUT}$. This is the case in our chosen scenario, since we have a single (LR-symmetric) intermediate scale.

For $m > 2$, the above system is under-determined. The general solution can then be written in terms of $m-2$ free parameters, which can be chosen to coincide with $m-2$ of the breaking scales. Nevertheless, in order to maintain the fixed order of the steps in the breaking chain, one needs to apply the constraint $\Delta t_i > 0$ on the scales, for all $i=1,\dots,m$. This condition reduces the allowed range for the independent scales.

Therefore, for all the models obtained as described in Section~\ref{sec:modellandscape} we obtain a set of limits (or exact solutions) for all scales consistent with the unification of gauge couplings. 

It is worth mentioning again that we neglect two-loop contributions to the RGEs, as well as threshold corrections and $U(1)$ mixing effects, which are all roughly of the same order. As we perform a rough scan over a large model landscape where we neglect model details (for example, heavy states are integrated out at exactly the same scale, but there could be a sizeable hierarchy between different masses), these approximations are well justified for our analysis. 

%---------------------------------------------------------
\section{Phenomenological Constraints}
\label{sec:phenoconstraints}
%---------------------------------------------------------
Given the large number of generated models, we will attempt to exclude some of them by considering their phenomenological consequences. The only model information we consider is the set of representations and the predictions of their mass scales from successful unification. There are therefore only a few phenomenological constraints that we apply, which will be outlined below. The list is by far not exhaustive but it aims to illustrate the effect of experimental constraints on the particle content and the corresponding interplay with the condition of gauge unification. There are a host of other important measurements such as electroweak precision observables, electric dipole moments, the anomalous magnetic moment of the muon and quark flavour observables. Clearly these play an important role, especially for a low LR-symmetry scale or under the presence of exotic states at the SM scale. The same applies to direct collider searches which for example exclude LR scales below a few TeV.

\subsection{Neutrino Masses}
%---------------------------------------------------------

In the SM, neutrinos are predicted to be massless, which cannot be reconciled with the observed oscillations of neutrino flavours~\cite{Pontecorvo:1967fh, Fukuda:1998mi}. Therefore, new physics models need to provide a source for those masses, often via some type of seesaw mechanism~\cite{Mohapatra:1979ia, Schechter:1980gr}. If the theory contains a right-handed neutrino $\nu^c$, the neutrino mass matrix takes the form
\begin{equation}
M_\nu = \left(
\begin{array}{cc}
m_L & m_D^T \\
m_D & m_R
\end{array} 
\right),
\end{equation}
where $m_D = y_\nu v$ is the Dirac-type neutrino mass matrix, proportional to the SM Higgs VEV $v$, with operator $m_D (\bar{\nu}_L \nu^c)$. The entries $m_L$ and $m_R$ are Majorana-type mass matrices for the left and right-handed neutrinos with operators $m_L (\nu_L \nu_L)$ and $m_R (\nu^c \nu^c)$, respectively. The Majorana mass $m_L$ would violate the SM gauge symmetries and thus can only be obtained through the VEV of a left-handed triplet Higgs $\Delta$, $m_L = \lambda_\Delta v_\Delta$.

In the model setup chosen in this analysis, three generations of right-handed neutrinos are always present, and we assume that they live at the LR scale. Therefore, the type I seesaw mechanism is always implemented, whereas type II seesaw, triggered by the Majorana mass of the left-handed neutrino $m_L$, relies on the presence of a triplet Higgs $\Delta$. The representations of $\nu^c$ and the triplet Higgs $\Delta$ for the SM, LR and $SO(10)$ gauge groups are displayed in Table~\ref{tab:nmassreps}.
\begin{table}[t!]
 \centering
 \begin{tabular}{cccc}
  \hline
  & SM & LR & $SO(10)$ \\
  \hline
  $\nu^c$ & $\{\mathbf{1},\mathbf{1},~0\}$ & $\{\mathbf{1},\mathbf{1},\mathbf{2},~1\}$ & $\mathbf{16}$ \\
  $\Delta$ & $\{\mathbf{1},\mathbf{3},~1\}$ & $\{\mathbf{1},\mathbf{3},\mathbf{1},~2\}$ & $\mathbf{126}$ \\
  \hline
 \end{tabular}
 \caption{Representations containing a right-handed neutrino $\nu^c$ and left-handed triplet Higgs $\Delta$.}
 \label{tab:nmassreps}
\end{table}

The current experimental bound on the mass of the neutrinos is $m_\nu^{\text{exp}} = \sum m_{\nu_i} \lesssim 0.3$ eV~\cite{Ade:2015xua}, with the lower limit given by the atmospheric mass splitting $\sqrt{\Delta m^2_\text{atm}} \approx 0.05$~eV for normally ordered neutrinos. Since the VEV of $\Delta$ is given by $v_\Delta \sim v^2/v_{LR}$ \cite{Mohapatra:1980yp}, one can take the following conservative range of the masses
\begin{equation}
	0.16 < \frac{m_\nu}{m_\nu^{\text{exp}}} \approx 
	\left| \lambda_\Delta - y_\nu^2  \right| 
	\left(\frac{2\times 10^{14} \text{ GeV}}{m_{LR}}\right) < 1.
	\label{neutrinomasses}
\end{equation}
In order to generate the observed light neutrino masses, the LR scale should thus be of the order of $10^{14}$~GeV, along the well known seesaw argument, with couplings $\lambda_\Delta, y_\nu = \mathcal{O}(1)$. Within the context of GUTs, neutrino masses may also be generated at the loop level under the presence of heavy leptoquarks \cite{Dorsner:2017wwn}. We here omit this possibility.

\subsection{Proton decay}
Because of the nature of GUTs, there are always exotic particles that couple to both quarks and leptons and could potentially mediate processes that violate baryon ($\mathcal{B}$) and/or lepton ($\mathcal{L}$) number~\cite{GellMann:1980vs}. SM interactions preserve both $\mathcal{B}$ and $\mathcal{L}$ perturbatively; baryon and lepton number violation is then introduced via non-renormalisable higher dimensional operators which can be probed by searching for very rare decays. The most important processes in GUTs that present this violation are proton decay, neutron-antineutron oscillations and lepton flavour violating processes. In addition, the total lepton number violating neutrinoless double beta decay is expected to occur if the light neutrinos are of Majorana nature, as discussed above.

The main decay mode of protons is $p\to e^+ \pi^0$, typically from dimension-6 operators, which could be mediated by a scalar or a gauge boson and suppressed by $M_X^{-2}$, the mass of the mediator. The $\Delta \mathcal{L} \neq 0$ and $\Delta \mathcal{B} \neq 0$ dimension 6 operators in the SM model are shown in Table \ref{tab:pd}~\cite{Senjanovic:2009kr}.
\begin{table}[t!]
 \centering
 \begin{tabular}{c|c}
   \hline
   $\mathcal{O}_6$ (gauge) & $\mathcal{O}_6$ (scalar)\\
   \hline
   $(\bar{Q}\sigma_\mu u^c) (\bar{L}\sigma^\mu d^c)\,,(\bar Q \sigma_\mu u^c) (\bar Q \sigma_\mu e^c) $ & $(Q Q) (Q L)\,, (Q Q) (\bar u^c \bar e^c)$ \\
   $(\bar{Q}\sigma_\mu d^c) (\bar{L}\sigma^\mu u^c)$ & $(u^c d^c) (\bar Q \bar L)\,, (u^c d^c) (u^c e^c)$ \\
   \hline
 \end{tabular}
 \caption{Dimension-6 operators that contribute to proton decay, mediated by gauge and scalar bosons.}
 \label{tab:pd}
\end{table}

For the operators in Table~\ref{tab:pd}, the SM representations of heavy bosons (gauge or scalar) that UV-complete them at tree level are quite constrained, cf. Table~\ref{tab:pdreps}. Since contributions to proton decay can arise at any step of the breaking chain, the completions of the SM reps in the LR and $SO(10)$ groups (with dimensions of representations lower than 200) are also provided.
\begin{table}[t!]
 \centering
 \begin{tabular}{r|ccc}
   \hline
          & SM & LR & $SO(10)$ \\
   \hline
   gauge & $\{\mathbf{3},\mathbf{2},\tminus\tfrac{5}{6}\}$, $\{\mathbf{3},\mathbf{2},~\tfrac{1}{6}\}$ &  $\{\mathbf{3},\mathbf{2},\mathbf{2},\tminus\tfrac{2}{3}\}$ & $\mathbf{45}$, $\mathbf{54}$\\
   \hline
   scalar & $\{\mathbf{3},\mathbf{3},\tminus\tfrac{1}{3}\}$, $\{\mathbf{3},\mathbf{1},\tminus\tfrac{1}{3}\}$  &  $\{\mathbf{3},\mathbf{3},\mathbf{1},\tminus\tfrac{2}{3}\}$, $\{\mathbf{3},\mathbf{1},\mathbf{3},\tminus\tfrac{2}{3}\}$   & $\mathbf{10}$, $\mathbf{120}$\\
   & & $\{\mathbf{3},\mathbf{1},\mathbf{1},\tminus\tfrac{2}{3}\}$ &  $\mathbf{126}$, $\overline{\mathbf{126}}$ \\
   \hline
 \end{tabular}
 \caption{Possible representations of gauge and scalar proton decay mediators and their completions in the LR and $SO(10)$ groups. Conjugates of these representations are also considered.}
 \label{tab:pdreps}
\end{table}

In a simple approximation, we estimate the proton decay half-life as~\cite{Olive:2014kda}.
\begin{equation}
 \tau_{p}^{\text{gauge}} \approx \frac{1}{\alpha_X^2}\frac{M_X^4}{m_p^5}, \quad \tau_{p}^{\text{scalar}} \approx \frac{(4\pi)^2}{\bar\lambda^4} \frac{M_X^4}{m_p^5},
\end{equation}
for gauge and scalar mediators respectively, where $\alpha_X$ is the gauge coupling fine structure constant at the unification scale $M_X$, $\bar \lambda$ is an average of the Yukawa-type couplings involved for a scalar mediator and $m_p$ is the proton mass. The experimental bound $\tau_p^{\text{exp}}$ on the proton decay half-life is given by
\begin{equation}
   \tau_p^\text{\text{exp}} > 1.29\times 10^{34} \text{ yr},
   \label{protondecayexp}
\end{equation}
determined by the Super-Kamiokande collaboration \cite{Nishino:2012bnw}. Hence, we can assess the proton decay contribution from gauge and scalar mediators as
\begin{align}
 \frac{\tau_{p}^\text{\text{gauge}}}{\tau_p^\text{\text{exp}}} \approx \frac{1}{\alpha^2_X}\left(\frac{M_{X}}{2.6\times10^{16}\text{ GeV}}\right)^4, \quad
 \frac{\tau_{p}^\text{\text{scalar}}}{\tau_p^\text{\text{exp}}} \approx \frac{1}{\bar\lambda^4}\left(\frac{M_{X}}{7.3\times10^{15}\text{ GeV}}\right)^4.
 \label{protondecay}
\end{align}

There are many other decay channels for protons, such as $p \to \mu^+\pi^0$, $p\to\nu K^+$ or $p\to e^+ K^0$ (see \cite{Nath:2006ut} for a full review). Most of these are mediated by the same operators and mediators as above, but their experimental lower bound is considerably lower than the main decay channel (by an order of magnitude or even more). Therefore, we focus exclusively on $p \to e^+ \pi^0$ and reasonably assume that if a model avoids this proton decay bound, it will also avoid all others.

\subsection{Neutron-antineutron oscillations}

\begin{figure}[ht]
 \centering
 \includegraphics[width=0.3\textwidth]{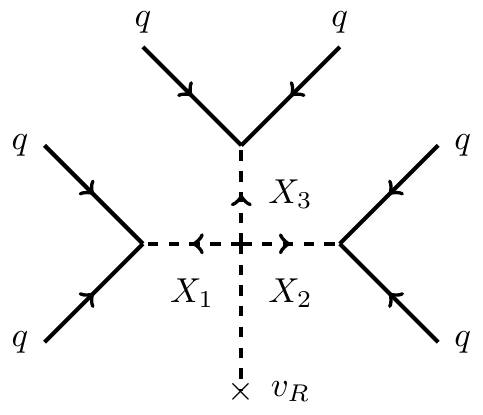}
 \caption{Schematic diagram leading to $n-\bar{n}$ oscillations, with three scalar mediators $X_i$.}
 \label{fig:nnoscillations}
\end{figure}

Similarly to the case of proton decay, high dimensional operators may induce $\mathcal{B}$ violating interactions which mix the neutron and antineutron mass states, cf. Figure~\ref{fig:nnoscillations}. At the quark level, such an $n-\bar{n}$ oscillation, violating $\Delta \mathcal{B} = 2$, is mediated by 9-dimensional operators of the type shown in Table~\ref{tab:nnoperators} ~\cite{Kuo:1980ew, Ozer:1982qh, Arnold:2012sd, Babu:2013jba}.
\begin{table}[t!]
 \centering
 \begin{tabular}{cc}
 \hline
 $\mathcal{O}_9$ (scalar) & $\mathcal{O}_9$ (scalar+vector) \\
 \hline
 $(Q Q) (Q Q) (\bar{d}^c \bar{d}^c)$ & $(\bar{Q}\sigma_\mu d^c) (\bar{Q}\sigma_\nu d^c) (\bar{Q} \bar{Q})$  \\
 $(Q Q) (\bar{d}^c \bar{d}^c) (\bar{u}^c \bar{d}^c)$ & $(\bar{Q}\sigma_\mu d^c) (\bar{Q}\sigma_\nu d^c) (u^c d^c)$ \\
 $(u^c u^c) (d^c d^c) (d^c d^c)$ & $(\bar{Q}\sigma_\mu d^c) (\bar{Q}\sigma_\nu u^c) (d^c d^c)$ \\ 
 $(u^c d^c) (u^c d^c) (d^c d^c)$ &  \\
 \hline
 \end{tabular}
 \caption{Dimension-9 operators contributing to $n-\bar n$ oscillations, involving three scalar currents (left), and one scalar plus two vector currents (right).}
 \label{tab:nnoperators}
\end{table}

As shown in Figure~\ref{fig:nnoscillations}, the coupling of the internal fields requires the insertion of the LR symmetry breaking VEV $v_R$ of the right-handed triplet Higgs.  All operators thus have coefficients of the order $v_R/M_X^6$, where $M_X$ is the average scale of the mediator masses $M_{X_i}$. This is due to the fact that $n-\bar{n}$ oscillations violate baryon number by two units, with no violation of lepton number, but $B-L$ is an exact symmetry at the LR and $SO(10)$ scales. Consequently, the only contributions at the LR and $SO(10)$ scales include an external scalar leg, that of the LR symmetry breaking field $\Delta_R \equiv \{\mathbf{1},\mathbf{1},\mathbf{3},~2\}_{LR} \in \mathbf{126}_{SO(10)}$~\cite{Mohapatra:1996pu}. This has the additional consequence of not allowing diagrams with three internal gauge bosons, because their coupling with $\Delta_R$ would violate Lorentz invariance. 

As in the case of proton decay, only certain representations have the necessary charges to UV-complete the operators in Table~\ref{tab:nnoperators} at the tree level. Because there are three mediators in this case, the representations must appear in certain $\mathcal{B}$ number violating combinations. We list those involving only scalar mediators in Table~\ref{tab:nnreps}.
\begin{table}[t!]
 \centering
 \small
 \begin{tabular}{c|c|c}
  \hline
  SM & LR & $SO(10)$ \\
  \setlength\tabcolsep{1.5pt}
  \begin{tabular}[t]{ccc}
  $X_1$ & $X_2$ & $X_3$ \\
  \hline 
  $\{\mathbf{6},\mathbf{3},~\tfrac{1}{3}\}$ & $\{\mathbf{6},\mathbf{3},~\tfrac{1}{3}\}$ & $\{\mathbf{6},\mathbf{1},\tminus\tfrac{2}{3}\}$ \\
  $\{\mathbf{6},\mathbf{3},~\tfrac{1}{3}\}$ & $\{\mathbf{\bar{3}},\mathbf{3},~\tfrac{1}{3}\}$ & $\{\mathbf{\bar{3}},\mathbf{1},\tminus\tfrac{2}{3}\}$ \\
  $\{\mathbf{6},\mathbf{1},~\tfrac{1}{3}\}$ & $\{\mathbf{6},\mathbf{1},~\tfrac{1}{3}\}$ & $\{\mathbf{6},\mathbf{1},\tminus\tfrac{2}{3}\}$ \\
  $\{\mathbf{6},\mathbf{1},~\tfrac{1}{3}\}$ & $\{\mathbf{\bar{3}},\mathbf{1},~\tfrac{1}{3}\}$ & $\{\mathbf{\bar{3}},\mathbf{1},\tminus\tfrac{2}{3}\}$ \\
  $\{\mathbf{6},\mathbf{1},~\tfrac{4}{3}\}$ & $\{\mathbf{6},\mathbf{1},\tminus\tfrac{2}{3}\}$ & $\{\mathbf{6},\mathbf{1},\tminus\tfrac{2}{3}\}$ \\
  $\{\mathbf{6},\mathbf{1},~\tfrac{4}{3}\}$ & $\{\mathbf{\bar{3}},\mathbf{1},\tminus\tfrac{2}{3}\}$ & $\{\mathbf{\bar{3}},\mathbf{1},\tminus\tfrac{2}{3}\}$ \\
  $\{\mathbf{\bar{3}},\mathbf{3},~\tfrac{1}{3}\}$ & $\{\mathbf{\bar{3}},\mathbf{3},~\tfrac{1}{3}\}$ & $\{\mathbf{6},\mathbf{1},\tminus\tfrac{2}{3}\}$ \\
  $\{\mathbf{\bar{3}},\mathbf{3},~\tfrac{1}{3}\}$ & $\{\mathbf{\bar{3}},\mathbf{3},~\tfrac{1}{3}\}$ & $\{\mathbf{\bar{3}},\mathbf{1},\tminus\tfrac{2}{3}\}$ \\
  $\{\mathbf{\bar{3}},\mathbf{1},~\tfrac{1}{3}\}$ & $\{\mathbf{\bar{3}},\mathbf{1},~\tfrac{1}{3}\}$ & $\{\mathbf{6},\mathbf{1},\tminus\tfrac{2}{3}\}$ \\
  $\{\mathbf{\bar{3}},\mathbf{1},~\tfrac{1}{3}\}$ & $\{\mathbf{\bar{3}},\mathbf{1},~\tfrac{1}{3}\}$ & $\{\mathbf{\bar{3}},\mathbf{1},\tminus\tfrac{2}{3}\}$ 
 \end{tabular} &
 \setlength\tabcolsep{1.5pt}
 \begin{tabular}[t]{ccc}
  $X_1$ & $X_2$ & $X_3$ \\
  \hline 
  $\{\mathbf{6},\mathbf{3},\mathbf{1},\tfrac{2}{3}\}$ & $\{\mathbf{6},\mathbf{3},\mathbf{1},\tfrac{2}{3}\}$ & $\{\mathbf{6},\mathbf{1},\mathbf{3},\tfrac{2}{3}\}$ \\
  $\{\mathbf{6},\mathbf{3},\mathbf{1},\tfrac{2}{3}\}$ & $\{\mathbf{\bar{3}},\mathbf{1},\mathbf{3},\tfrac{2}{3}\}$ & $\{\mathbf{\bar{3}},\mathbf{3},\mathbf{1},\tfrac{2}{3}\}$ \\
  $\{\mathbf{6},\mathbf{1},\mathbf{3},\tfrac{2}{3}\}$ & $\{\mathbf{6},\mathbf{1},\mathbf{3},\tfrac{2}{3}\}$ & $\{\mathbf{6},\mathbf{1},\mathbf{3},\tfrac{2}{3}\}$ \\
  $\{\mathbf{6},\mathbf{1},\mathbf{3},\tfrac{2}{3}\}$ & $\{\mathbf{6},\mathbf{1},\mathbf{1},\tfrac{2}{3}\}$ & $\{\mathbf{6},\mathbf{1},\mathbf{1},\tfrac{2}{3}\}$ \\
  $\{\mathbf{6},\mathbf{1},\mathbf{3},\tfrac{2}{3}\}$ & $\{\mathbf{\bar{3}},\mathbf{1},\mathbf{3},\tfrac{2}{3}\}$ & $\{\mathbf{\bar{3}},\mathbf{1},\mathbf{3},\tfrac{2}{3}\}$ \\
  $\{\mathbf{6},\mathbf{1},\mathbf{3},\tfrac{2}{3}\}$ & $\{\mathbf{\bar{3}},\mathbf{3},\mathbf{1},\tfrac{2}{3}\}$ & $\{\mathbf{\bar{3}},\mathbf{3},\mathbf{1},\tfrac{2}{3}\}$ \\
  $\{\mathbf{6},\mathbf{1},\mathbf{3},\tfrac{2}{3}\}$ & $\{\mathbf{\bar{3}},\mathbf{1},\mathbf{1},\tfrac{2}{3}\}$ & $\{\mathbf{\bar{3}},\mathbf{1},\mathbf{1},\tfrac{2}{3}\}$ \\
  $\{\mathbf{6},\mathbf{1},\mathbf{1},\tfrac{2}{3}\}$ & $\{\mathbf{\bar{3}},\mathbf{1},\mathbf{3},\tfrac{2}{3}\}$ & $\{\mathbf{\bar{3}},\mathbf{1},\mathbf{1},\tfrac{2}{3}\}$ \\
  $\{\mathbf{\bar{3}},\mathbf{1},\mathbf{3},\tfrac{2}{3}\}$ & $\{\mathbf{\bar{3}},\mathbf{1},\mathbf{3},\tfrac{2}{3}\}$ & $\{\mathbf{\bar{3}},\mathbf{1},\mathbf{3},\tfrac{2}{3}\}$ \\
  $\{\mathbf{\bar{3}},\mathbf{1},\mathbf{3},\tfrac{2}{3}\}$ & $\{\mathbf{\bar{3}},\mathbf{3},\mathbf{1},\tfrac{2}{3}\}$ & $\{\mathbf{\bar{3}},\mathbf{3},\mathbf{1},\tfrac{2}{3}\}$  \\
  $\{\mathbf{\bar{3}},\mathbf{1},\mathbf{3},\tfrac{2}{3}\}$ & $\{\mathbf{\bar{3}},\mathbf{1},\mathbf{1},\tfrac{2}{3}\}$ & $\{\mathbf{\bar{3}},\mathbf{1},\mathbf{1},\tfrac{2}{3}\}$ 
 \end{tabular} &
 \setlength\tabcolsep{1.5pt}
 \begin{tabular}[t]{ccc}
  $X_1$ & $X_2$ & $X_3$ \\
  \hline 
  $\mathbf{10}$ & $\mathbf{10}$ & $\mathbf{120}$ \\
  $\mathbf{10}$ & $\mathbf{10}$ & $\overline{\mathbf{126}}$ \\
  $\mathbf{10}$ & $\mathbf{120}$ & $\mathbf{120}$ \\
  $\mathbf{10}$ & $\mathbf{120}$ & $\overline{\mathbf{126}}$ \\
  $\mathbf{120}$ & $\mathbf{120}$ & $\mathbf{120}$ \\
  $\mathbf{120}$ & $\mathbf{120}$ & $\overline{\mathbf{126}}$ \\
  $\mathbf{120}$ & $\overline{\mathbf{126}}$ & $\overline{\mathbf{126}}$ \\
 \end{tabular} \\
 \hline
 \end{tabular}
 \caption{Combination of scalar representations mediating $n-\bar{n}$ oscillations at the tree level, cf. Figure~\ref{fig:nnoscillations}, for the SM, LR and $SO(10)$ scales. The equivalent combinations with conjugate representations are also considered.}
 \label{tab:nnreps}
\end{table}

The contributions to $n-\bar{n}$ oscillations coming from gauge bosons can be neglected at this stage. This is due to the fact that all gauge representations we consider at the SM and LR scales are diagonal in the gauge groups and in particular have $Y=0$ and $B-L = 0$, respectively, which disallows the construction of the relevant operators. At the $SO(10)$ level, however, the operator can be constructed using the gauge bosons from the $\mathbf{45}$ representation, yet this contribution is negligible since it is suppressed by $M_X^{-6}$ and the $SO(10)$ scale is expected to be of the order of $10^{16}$~GeV or higher.

In general, each of the internal mediators $X_{1,2,3}$ can live at any scale $M_{X_{1,2,3}}$. Therefore, the contribution to the matrix element for the transition $n \to\bar{n}$ can be estimated as \cite{Babu:2013jba}
\begin{equation}
 \frac{\delta m}{\delta m^{\text{exp}}} \approx \kappa \bar\lambda^3 \frac{(4.7 \times 10^{5}\text{ GeV})^5~M_{LR} }{M_{X_1}^2 M_{X_2}^2 M_{X_3}^2},
 \label{nnmatrixelement}
\end{equation}
with $\kappa$ the 4-scalar coupling and $\bar\lambda$ the average of the trilinear couplings of the $X_i$ to quarks in Figure~\ref{fig:nnoscillations}.

Though the process of neutron-antineutron oscillation is heavily suppressed by five powers of a heavy scale, its experimental limit is still very severe, $\delta m^{\text{exp}} < 2.81 \times 10^{-33}$~GeV ($\tau_{n\bar n} < 2.7 \times 10^8 $~s)~\cite{Abe:2011ky}, and thus can be relevant if any or all of the mediators $X_i$ appear at lower scales.

\subsection{Lepton Flavour Violation}

Lepton flavour violation (LFV) is a particular case of lepton number violation, where the individual lepton number of a generation $\mathcal{L}_i$ is violated, while preserving the overall sum~\cite{Kuno:1999jp}. For example, the well studied process of muon decay, $\mu \to e\gamma$, has $\Delta\mathcal{L}_\mu = -1$ and $\Delta\mathcal{L}_e = 1$. Despite the presence of LFV processes in the SM, namely neutrino oscillation via the PMNS matrix~\cite{Pontecorvo:1959sn, Maki:1962mu}, there is no equivalent process for charged leptons. Therefore, the search for charged lepton flavour violation (CLFV) provides another highly stringent constraint on new physics models~\cite{Gluza:2016qqv}. 

The most commonly studied processes, also with the highest experimental sensitivity, are the rare muon decays $\mu \to e\gamma$, $\mu \to eee$ and $\mu - e$ conversion in nuclei. The current experimental limits for these processes are $\mathcal{B}(\mu\to e\gamma)_{\text{exp}} < 5.7 \times 10^{-13}$ \cite{Adam:2013mnn}, $\mathcal{B}(\mu\to eee)_{\text{exp}} < 1.0 \times 10^{-12}$ \cite{Bellgardt:1987du} and $\mathcal{B}(\mu N \to e N)_{\text{exp}} < 7 \times 10^{-13}$ \cite{Bertl:2006up}.

Beyond the SM, these processes are triggered by dimension-6 operators\footnote{In fact, diagrams with neutrinos and a SM $W$ boson in the loop are present in the SM with massive neutrinos but their contribution is heavily suppressed because of the tiny mass of the neutrinos.}. The photonic dipole operators for $\mu \to e\gamma$, which also contribute to $\mu\to eee$ and $\mu - e$ conversion via photon exchange, are listed on the left-hand side of Table~\ref{tab:mutoeoperators} \cite{Kuno:1999jp}. In addition, the processes $\mu\to eee$ and $\mu - e$ conversion in nuclei can also be mediated by four-fermion interactions and their corresponding effective operators are listed on the right-hand side in Table~\ref{tab:mutoeoperators} \cite{Kuno:1999jp}. 
\begin{table}[t!]
 \centering 
 \begin{tabular}{c}
  \hline
  $M_X^2 \mathcal{O}_{\mu\to e\gamma}$ \\
  \hline
  $m_{f} \mu \sigma_{\mu\nu} e^c F^{\mu\nu}$\\
  $m_{f} \mu^c \sigma_{\mu\nu} e F^{\mu\nu}$\\
  \hline \\ \\ \\ \\
 \end{tabular}\quad
 \begin{tabular}{cc}
  \hline
  $M_X^2 \mathcal{O}_{\mu\to eee}$ &  $M_X^2 \mathcal{O}_{\mu-e}$ \\
  \hline
  $(\mu~ e^c) (e~ e^c)$ & $(\mu~ e^c) (q~q^c)$\\
  $(\mu^c~ e) (e^c~ e)$ &  $(\mu^c~e) (q~q^c)$\\
  \hline
  $(\bar \mu \bar\sigma_\mu e) (\bar e \bar\sigma^\mu e)$ & $(\bar\mu \bar\sigma_\mu e) (\bar q \bar\sigma^\mu q)$\\
  $(\bar \mu^c \bar\sigma_\mu e^c) (\bar e \bar\sigma^\mu e)$ & $(\bar\mu^c \bar\sigma_\mu e^c) (\bar q \bar\sigma^\mu q)$ \\
  $(\bar \mu \bar\sigma_\mu e) (\bar e^c \bar\sigma^\mu e^c)$ & $(\mu \sigma_{\mu\nu} e^c) (q^c \sigma^{\mu\nu} q)$ \\
  $(\bar \mu ^c \bar\sigma_\mu e^c) (\bar e^c \bar\sigma^\mu e^c)$ & $(\mu^c \sigma_{\mu\nu} e) ( q^c \sigma^{\mu\nu} q)$  \\
  \hline
 \end{tabular}
 \caption{Effective operators for $\mu\to e\gamma$, $\mu\to eee$ and $\mu - e$ conversion. The table on the left shows the dipole operator contributing to all three processes while the table on the right shows the contribution of four fermion diagrams to $\mu \to eee$ and $\mu - e$ with scalar (top) and gauge (bottom) mediators. The SM-invariant operators are of dimension-6 and are thus suppressed by $1/M_X^2$ where $M_X$ is the typical scale of new physics generating the operator. The mass $m_f$ is that of the muon or a heavier fermion in the loop generating the dipole operators.} 
 \label{tab:mutoeoperators}
\end{table}

With the LR symmetry group at an intermediate and possibly low scale, gauge and gauge-breaking Higgs contributions to the above LFV processes are of special importance. The radiative decay $\mu\to e\gamma$ is dominantly mediated by a right-handed gauge boson $W_R \equiv \{\mathbf{1},\mathbf{1},\mathbf{3},~0\}$ and heavy right-handed neutrinos in the loop, as is $\mu-e$ conversion in nuclei through box diagrams generated by the four fermion operators on the right-hand side of Table \ref{tab:mutoeoperators}¸. With the presence of a right-handed  triplet Higgs $\Delta_R \equiv \{\mathbf{1},\mathbf{1},\mathbf{3},~2\}$ in all scenarios we consider, $\mu\to eee$ is triggered at tree level via the exchange of the doubly-charged Higgs triplet. While generally depending on many parameters, especially through the flavour structure in the right-handed lepton sector, these contributions can be very well approximated assuming consummate mass scales among the relevant heavy particles in the LR symmetric model, $M_{LR} \equiv m_{N_i} \approx m_{W_R} \approx m_{\Delta_R}$~\cite{Cirigliano:2004mv}, as
\begin{align}
\label{eq:BrmuegammaSimplified}
\frac{B(\mu\to e\gamma)}{B(\mu\to e\gamma)_{\text{exp}}}
&\approx |g_{e\mu}|^2 \left(\frac{g_R}{g_L}\right)^4  \left(\frac{2.3\times 10^4 \text{ GeV}}{M_{LR}}\right)^4, \\
\label{eq:BrmueSimplified}
\frac{B(\mu - e)}{B(\mu - e)_{\text{exp}}}
&\approx |g_{e\mu}|^2 \left(\frac{g_R}{g_L}\right)^4 
\left(\frac{2.1\times 10^4 \text{ GeV}}{M_{LR}}\right)^4, \\
\label{eq:BrmueeeSimplified}
\frac{B(\mu\to eee)}{B(\mu\to eee)_{\text{exp}}}
&\approx |g_{e\mu}|^2 \left(\frac{9.1\times 10^4 \text{ GeV}}{M_{LR}}\right)^4.
\end{align}
Here, $g_{e\mu}$ is an effective lepton-gauge boson coupling in (quasi-)\-manifest LR symmetry,
\begin{align}
\label{eq:LFVCouplings}
g_{e\mu} &= 
\sum_{n=1}^3 V^*_{en} V^{\phantom{\dagger}}_{\mu n}
\left(\frac{m_{N_n}}{m_{W_R}}\right)^2.
\end{align}
The $3\times 3$ matrix $V$ is the mixing matrix of the right-handed charged current, i.e. the equivalent of the PMNS matrix for heavy right-handed neutrinos.

In addition to the contributions mediated by $W_R$ and $\Delta_R$, the dipole operators on the left-hand side of Table~\ref{tab:mutoeoperators} can be mediated by a exotic scalar/vector boson and a fermion in the loop. As already noted, these diagrams contribute to all LFV processes, even to $\mu \to eee$ and $\mu - e$ conversion, where the emitted photon acts as the internal mediator. All such diagrams require a mass insertion on the external or internal fermion line.
\begin{table}[t!]
 \centering 
 \begin{tabular}[t]{ccc}
  \hline
  SM & LR & $SO(10)$ \\
  \hline   
  {\color{gray} $\{\mathbf{1},\mathbf{2},~\tfrac{1}{2}\}_N$}, $\{\mathbf{1},\mathbf{1},~1\}_N$, & {\color{gray}$\{\mathbf{1},\mathbf{2},\mathbf{2},~0\}_N$}, $\{\mathbf{1},\mathbf{1},\mathbf{1},~2\}_N$, &  $\mathbf{10}$, $\mathbf{120}$, $\mathbf{126}$, $\mathbf{\overline{126}}$\\
  $\{\mathbf{3},\mathbf{2},~\tfrac{7}{6}\}_t$, $\{\mathbf{3},\mathbf{1},\tminus\tfrac{1}{3}\}_t$ & $\{\mathbf{3},\mathbf{2},\mathbf{2},~\tfrac{4}{3}\}_t$, $\{\mathbf{3},\mathbf{1},\mathbf{1},\tminus\tfrac{2}{3}\}_t$ & \\ 
  \hline
 \end{tabular}
 \caption{Possible representations for the mediating scalar field $X$ in the loop for $\mu\to e\gamma$, at the SM, LR and $SO(10)$ scales. The subscript indicates the associated fermion inside the loop. Conjugates of these representations are also considered. The greyed out representations correspond to the SM Higgs which we assume are not able to mediate LFV currents.}
 \label{tab:mutoegammareps}
\end{table}
As constructed, our models do not contain any exotic fermions and thus the fermion $f$ in the loop generating the dipole operator is a SM fermion. We are implicitly assuming that the mediators have all possible couplings to SM fermions, including flavour violating couplings. Along with allowing the LFV contribution to happen, this means that all SM fermions are allowed inside the loop, provided the gauge symmetry is not violated. Consequently, the scalar or gauge boson $X$ must be in a representation allowed by gauge invariance. The available representations for $X$ for $\mu \to e\gamma$ are shown in Table~\ref{tab:mutoegammareps} where, for the sake of completion, we have included as well the representations allowed for equivalent diagrams at the LR and $SO(10)$ scales.

The representations in Table~\ref{tab:mutoegammareps} include a subscript if their contribution is dependent on a particular fermion running in the loop, or no subscript if they contribute regardless of it. It is worth noting that we have not listed any gauge boson representations in Table~\ref{tab:mutoegammareps}. This is due to the fact that we do not consider exotic non-diagonal gauge bosons, but only the minimal set at each energy scale, and these do not mix flavours. The only possible contribution from SM gauge bosons to LFV is actually negligible, because it relies on neutrino mixing and it is proportional to the neutrino mass \cite{Deppisch:2012vj}.

With these assumptions, we estimate the contribution of the dipole operator to the LFV observables as \cite{Raidal:1997hq, Kuno:1999jp, Cirigliano:2004mv, Benbrik:2010cf, Altarelli:2010gt, Deppisch:2012vj, Chang:2016zll, FileviezPerez:2017zwm}
{\small 
\begin{align}
 \notag \frac{B(\mu\to e\gamma)}{B(\mu\to e\gamma)_{\text{exp}}} &\approx |\overline{\lambda_N^\mu \lambda_N^e}|^2 \left(\frac{3.0\times10^{4}\text{ GeV}}{M_X}\right)^4,\,\, |\lambda_t^\mu \lambda_t^e|^2 \left(\frac{2.2\times 10^4\text{ GeV}}{M_X} \right)^4, \\ 
 \notag \frac{B(\mu \to eee)^{\text{phot}}}{B(\mu\to eee)_{\text{exp}}} &\approx |\overline{\lambda_N^\mu \lambda_N^e}|^2 \left(\frac{7.3\times 10^{3} \text{ GeV}}{M_X}\right)^4,\,\, |\lambda_t^\mu \lambda_t^e|^2 \left(\frac{6.0\times 10^3\text{ GeV}}{M_X} \right)^4, \\ 
 \frac{B(\mu - e)^{\text{phot}}}{B(\mu - e)_{\text{exp}}} &\approx 
 |\overline{\lambda_N^\mu \lambda_N^e}|^2 \left(\frac{6.9\times 10^{3}\text{ GeV}}{M_X}\right)^4,\,\, 
 |\lambda_t^\mu \lambda_t^e|^2 \left(\frac{5.0\times 10^3\text{ GeV}}{M_X} \right)^4.
\end{align}
}%
Here, $\lambda_t^l$ denotes the coupling between the external lepton ($l=e,\mu$), the exotic scalar $X$ and the top in the loop, where we neglect the potential contributions of lighter quarks. The coupling $\lambda_N^l$ is the corresponding equivalent for the heavy neutrinos in the loop, where the overline indicates an average over the three neutrino generations.
\begin{table}[t!]
 \centering
 \begin{tabular}{ccc}
  \hline
  SM & LR & $SO(10)$ \\
  \hline
  $\{\mathbf{1},\mathbf{3},~1\}$, $\{\mathbf{1},\mathbf{1},~1\}$, & $\{\mathbf{1},\mathbf{3},\mathbf{1},~2\}$, $\{\mathbf{1},\mathbf{1},\mathbf{1},~2\}$, & $\mathbf{10}$, $\mathbf{120}$ \\
  {\color{gray}$\{\mathbf{1},\mathbf{2},~\tfrac{1}{2}\}$}, $\{\mathbf{1},\mathbf{1},~2\}$ & {\color{gray}$\{\mathbf{1},\mathbf{1},\mathbf{3},~2\}$}, {\color{gray}$\{\mathbf{1},\mathbf{2},\mathbf{2},~0\}$} & $\mathbf{126}$, $\overline{\mathbf{126}}$ \\
  \hline
  $\{\mathbf{3},\mathbf{3},~\tfrac{1}{3}\}$, $\{\mathbf{3},\mathbf{1},~\tfrac{4}{3}\}$, & $\{\mathbf{3},\mathbf{3},\mathbf{1},\tminus\tfrac{2}{3}\}$, $\{\mathbf{3},\mathbf{1},\mathbf{1},\tminus\tfrac{2}{3}\}$, & $\mathbf{10}$, $\mathbf{120}$ \\
  $\{\mathbf{3},\mathbf{1},~\tfrac{1}{3}\}$, $\{\mathbf{3},\mathbf{2},~\tfrac{7}{6}\}$, & $\{\mathbf{3},\mathbf{1},\mathbf{3},\tminus\tfrac{2}{3}\}$, $\{\mathbf{3},\mathbf{2},\mathbf{2},~\tfrac{4}{3}\}$ & $\mathbf{126}$, $\overline{\mathbf{126}}$ \\
  $\{\mathbf{3},\mathbf{2},~\tfrac{1}{6}\}$ &{\color{gray}$\{\mathbf{1},\mathbf{2},\mathbf{2},~0\}$} & \\
  \hline
 \end{tabular}
 \caption{Possible representations for the mediating scalar field $X$ for $\mu\to eee$ (top) and $\mu \to e$ conversion in nuclei (bottom) at the SM, LR and SO(10) scales. Conjugates of these representations are also considered. Greyed out representations correspond to the SM Higgs, which does not mediate LFV, and the right-handed triplet, which was already considered in Eq. \eqref{eq:BrmuegammaSimplified} - \eqref{eq:BrmueeeSimplified} because it is always present in LR models.}
 \label{tab:mutoereps}
\end{table}

Similarly, we estimate the contributions to $\mu\to eee$ and $\mu-e$ conversion via the tree level exchange of exotic scalars. The four-fermion diagrams can be generated at the tree level by a scalar boson, with the effective operators shown in the upper rows of the right-hand side of Table~\ref{tab:mutoeoperators}, or a gauge boson, with operators in the bottom rows. The possible representations for the scalar mediator are shown in Table~\ref{tab:mutoereps}. The corresponding contributions are then estimated as
\begin{align}
\notag \frac{B(\mu \to eee)^{\text{\st{phot}}}}{B(\mu\to eee)_{\text{exp}}} &\approx |(\lambda\lambda')_X^{e\mu}|^2 \left(\frac{6.8\times 10^{4} \text{ GeV}}{M_X}\right)^4, \\ 
\frac{B(\mu - e)^{\text{\st{phot}}}}{B(\mu - e)_{\text{exp}}} 
&\approx |(\lambda\lambda')_X^{e\mu}|^2 \left(\frac{2.7\times 10^{4}\text{ GeV}}{M_X} \right)^4,
\end{align}
where $(\lambda\lambda')_X^{e\mu}$ is the product of Yukawa-like couplings between the exotic scalar $X$ and the relevant fermions leading to a $\mu\to e$ lepton flavour transition.

%---------------------------------------------------------
\section{Results}
\label{sec:results}
%---------------------------------------------------------

\begin{figure}[t!]
 \centering
 \includegraphics[width=0.49\textwidth]{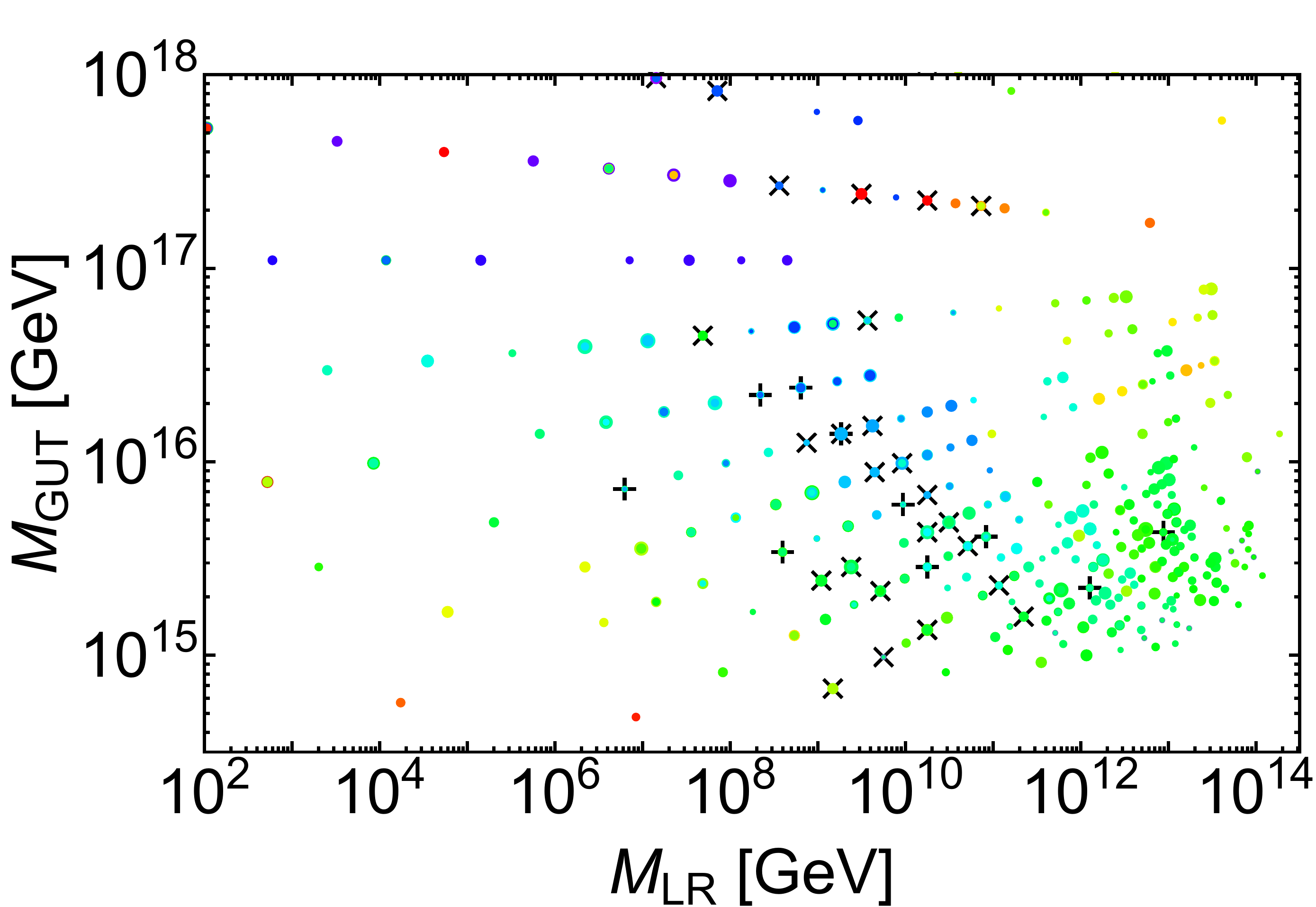}
 \includegraphics[width=0.49\textwidth]{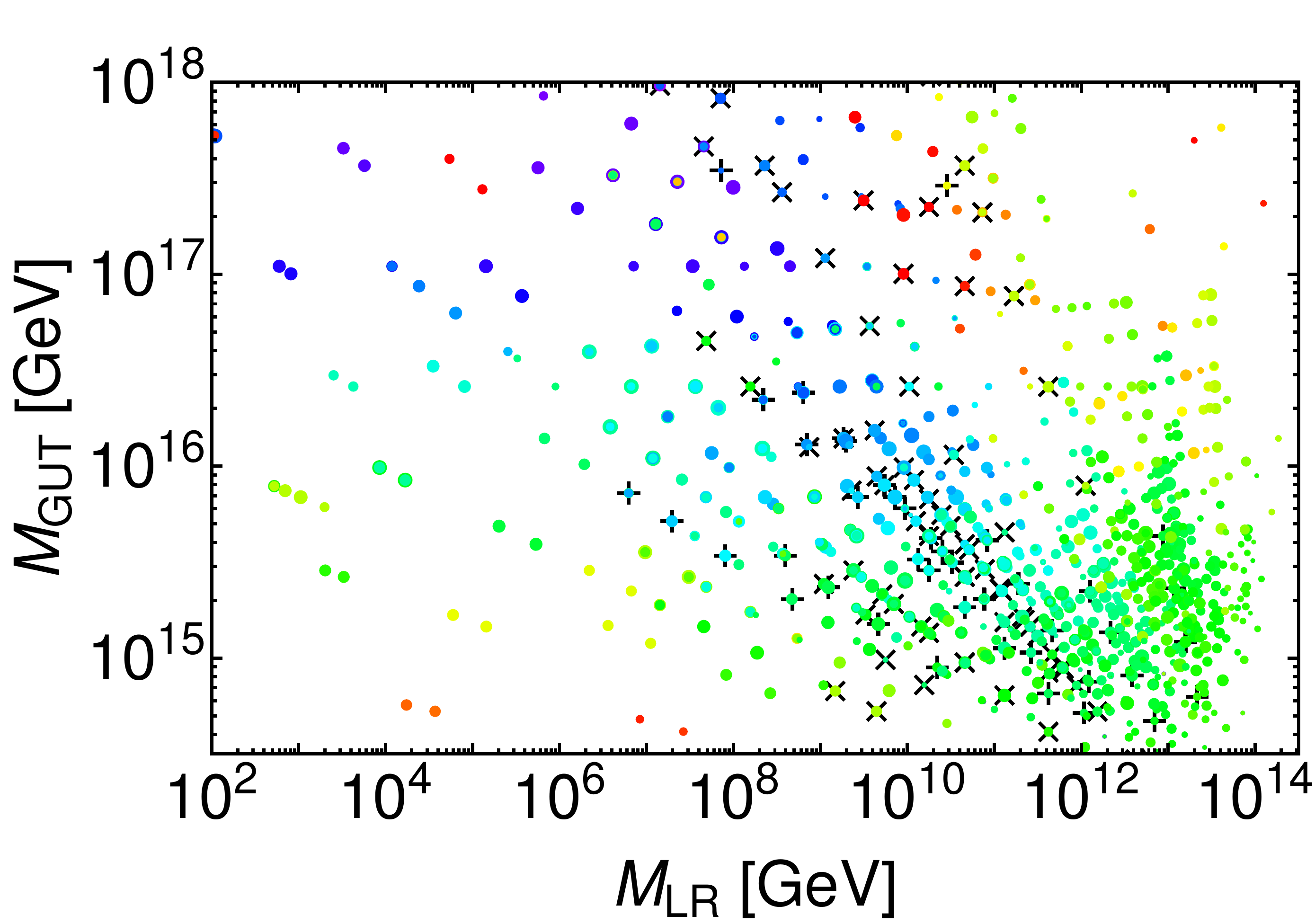}\\
 \includegraphics[width=0.87\textwidth]{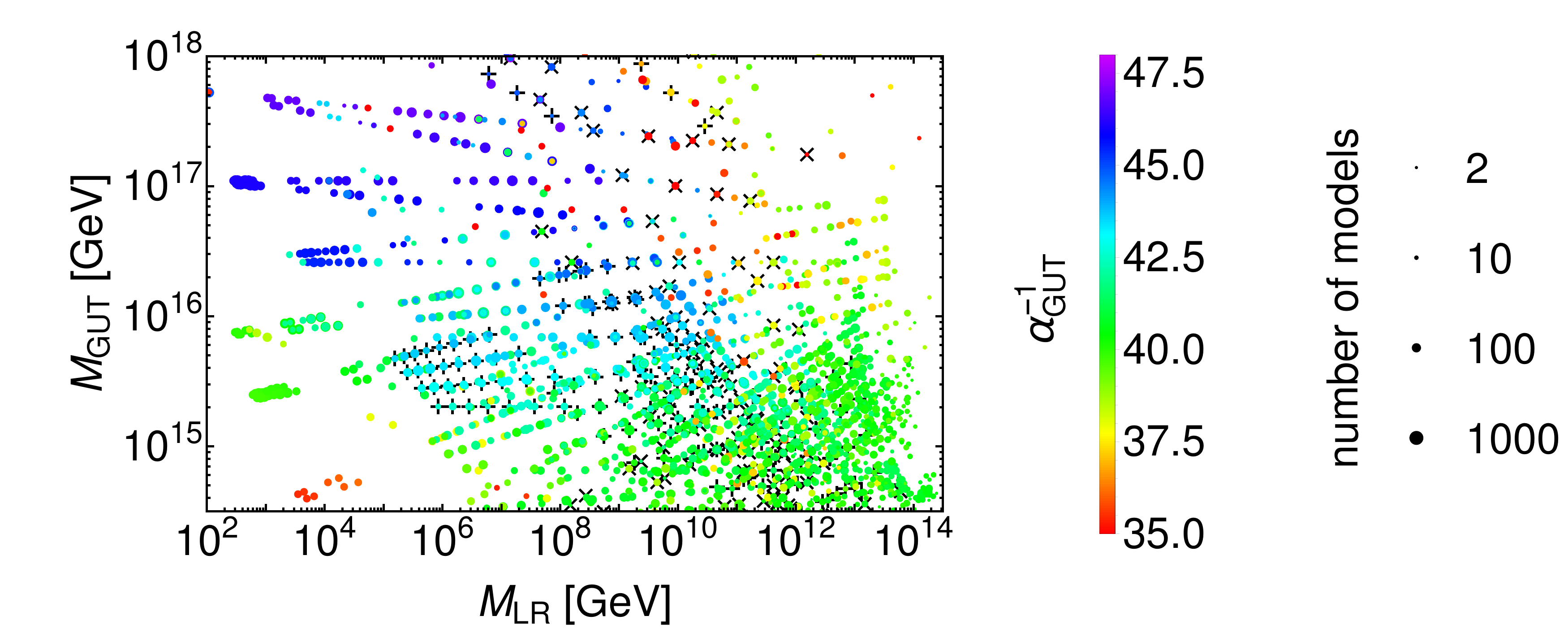}
 \caption{Solutions $(M_{GUT}, M_{LR}, \alpha_{GUT})$ to the unified RG running for models with no exotic fields at the SM scale (top left), only electroweak doublets and/or singlets at the SM scale (top right) and all models (bottom). The colours represent the value of $\alpha_\text{GUT}$ whereas the crosses identify models with manifest left-right symmetry ($\times$) and models that satisfy the CKM constraint ($+$). The point sizes indicate the multiplicity of individual models represented, as indicated in the legend.}
\label{fig:alpha}
\end{figure}

Within the large number of models generated by the algorithm described in Sections~\ref{sec:modellandscape} and \ref{sec:leftrightsymmetry}, there are many that include several exotic fields at the lowest energy scale. The presence of these additional exotics, even if they do not trigger any of the studied observables, might be controversial and may contradict current experimental limits not considered in this study. In particular this is the case for coloured exotics, which if present at low energies may have the undesired effect of modifying the QCD RGEs enough to contradict the well studied phenomenon of asymptotic freedom \cite{Ilisie:2012cc, Bethke:2011tr}, as well as current limits on QCD exotics \cite{Olsen:2014mea}. Because of this reason we will only study a subset of the total number of models, (i) those containing just the SM particle content plus additional singlets or (ii) those containing any number of singlets and $SU(2)$ doublets, aiming to include two-Higgs doublet models in the scan. Coloured exotics are allowed to exist at intermediated scales, i.e. LR scale, where they might contribute to observables such as proton decay.

In Figure \ref{fig:alpha} one can see the models scattered in the plane $M_{GUT} - M_{LR}$ for cases (i) and (ii) (top left and top right respectively), and the full set of models (bottom), where the colours represent the value of $\alpha_{GUT}$, ranging from $\approx 1/47$ (purple) to $\approx 1/33$ (red). The size of the dots in all figures represents the multiplicity of the individual models, since often there are several distinct models that have equal predictions for the energy scales and observables.

Most of the models obtained are not manifestly left-right symmetric, i.e. they are not invariant under the exchange $SU(2)_L \leftrightarrow SU(2)_R$. The small subset of models that show manifest LR symmetry are shown in Figure \ref{fig:alpha} with a cross ({$\times$) underneath the dot. These models typically show a preference for mid-range LR scale, $M_{LR} \sim 10^{8-11}$, most likely due to the increased number of representations present in the model.

As was anticipated above, in Section \ref{sec:constraints}, in order to reproduce the SM fermion masses, i.e. the CKM matrix, a model must contain a specific set of representations. For our LR model, these must be at least two copies of the scalar bidoublet $\{\mathbf{1},\mathbf{2},\mathbf{2},~0\}$ and one copy of a right-handed triplet $\{\mathbf{1},\mathbf{1},\mathbf{3},~0\}$. In Figure \ref{fig:alpha} we have highlighted models that satisfy this constraint with a plus sign ($+$) underneath. Though there does not seem to be a clear pattern of these models, there is a rather slight preference towards mid to large $M_{LR}$, but less so than the manifestly LR symmetric models above.

All three plots in Figure \ref{fig:alpha} show a higher concentration of models around $M_{GUT} \sim 10^{15}$ GeV and $M_{LR} \sim 10^{13}$ GeV. This is probably due to the fact that the SM gauge couplings naturally run towards each other around $10^{15}$ GeV, and a small number ($N \leq 5$) of exotic representations (some of them may be singlets) cannot easily bend the RGEs significantly away from that pattern. That said, there are still some models with a much larger GUT scale, even close to the Planck scale $M_P \sim 10^{18}$ GeV where gravitational corrections to the gauge couplings may have to be considered \cite{Robinson:2005fj}.

\begin{table}[h]
 \centering
 \begin{tabular}{rll}
  \hline
  Observable & Current & Future \\
  \hline
  $\tau^{\text{exp}}_p$ & $1.3 \times 10^{34}$ y & $1.3 \times 10^{35}$ y \cite{Abe:2015zbg}\\
  $\tau^{\text{exp}}_{n-\bar n}$ & $2.7 \times 10^8$ s & $6.8 \times 10^8$ s \cite{Phillips:2014fgb} \\
  $\mathcal{B}(\mu\to e\gamma)$ & $5.7 \times 10^{-13}$ & $4.0 \times 10^{14}$ \cite{Renga:2014xra} \\
  $\mathcal{B}(\mu\to eee)$ & $1.0 \times 10^{-12}$ & $1.0 \times 10^{-16}$ \cite{Perrevoort:2016nuv}\\
  $\mathcal{B}(\mu N\to eN)$ & $7.0 \times 10^{-13}$ & $6.0 \times 10^{-17}$ \cite{Morescalchi:2016uks} \\
  \hline
 \end{tabular}
 \caption{Current and future limits on the phenomenological constraints.}
 \label{tab:limits}
\end{table}

The next step is to constrain the set of models displayed in Figure \ref{fig:alpha} using the phenomenological observables from Section \ref{sec:phenoconstraints}. For completion we show in Table \ref{tab:limits} the current experimental limits for those observables, as well as the predicted limit for the next generation of experiments.

Figure \ref{fig:results-protondecay} shows one of the most constraining observables, proton decay, for models with only the SM particles (left) and the models with extra singlets and $SU(2)$ doublets (right). As before, the size of the dots in both figures represents the multiplicity of the individual models. The colours in the plot represent the potential contribution of exotic scalars fields in the given model to proton decay. Blue dots have no contributions whereas for the green, orange and red points, dangerous proton decay rates may be triggered if the corresponding couplings are of order $10^{-2} - 1$, $10^{-4} - 10^{-2}$ and $10^{-6} - 10^{-4}$, respectively. The shading and horizontal lines represent the contribution from gauge interactions. The most desaturated dots, below the solid line, are excluded with the current proton decay limits. The next level of saturation, below the dashed line, would be excluded assuming one order of magnitude increase on the experimental limit. Lastly, the high saturation points have no dangerous proton decay gauge contributions.

\begin{figure}[ht!]
 \centering
 \includegraphics[width=0.48\textwidth]{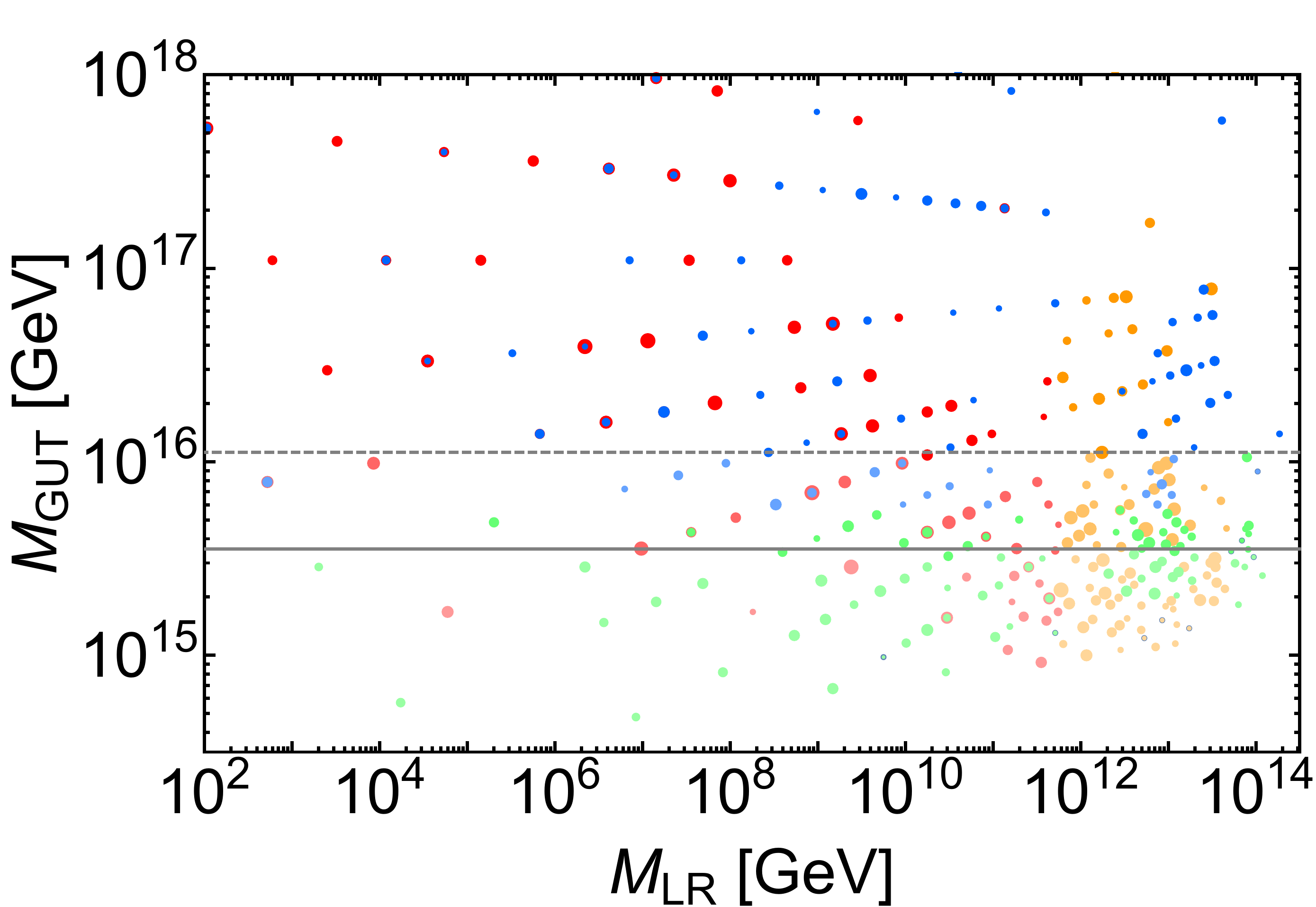}
 \includegraphics[width=0.48\textwidth]{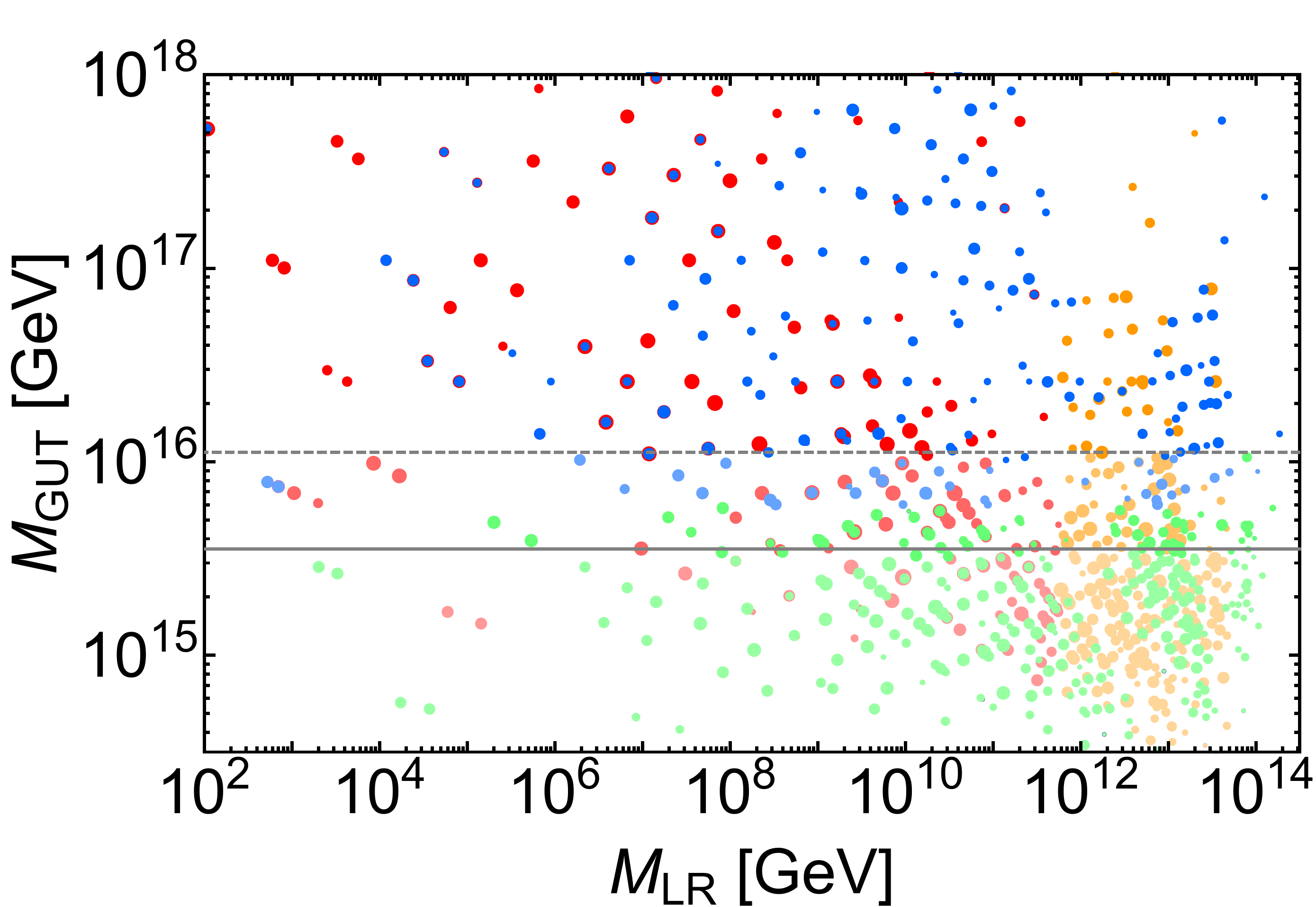}
 \caption{Effect of the proton decay constraint on the solutions $(M_{GUT}, M_{LR})$ of the unified RG running for models with no exotic fields at the SM scale (left) and with only electroweak doublets and/or singlets at the SM scale (right). The colours represent the potential contribution of exotic scalar fields in the given model to proton decay, while the shading and horizontal lines show the gauge contribution. The point sizes indicate the multiplicity of models.}
\label{fig:results-protondecay}
\end{figure}

As expected, the gauge contribution excludes values of the unification scale below $\sim 10^{15-16}$ GeV with the current experimental limit, and somewhat larger values with the projected future limit. The scalar contributions have an almost arbitrary pattern, but show a preference towards large values of $M_{GUT}$ and mid-range values of $M_{LR}$. Proton decay is significantly the biggest constraint on the value of the GUT scale, while other observables, such as neutron-antineutron oscillations provide a much weaker constraint and thus we will not show them on a separate plot. It can be noticed that there is a large number of scenarios designated with red points at low LR scales; these correspond to models with coloured exotics at the LR scale which trigger fast proton decay even for small values of the couplings.

Conversely, there is a number of observables that constrain the value of $M_{LR}$, leaving $M_{GUT}$ almost unscathed. We here discuss lepton flavour violating decays, as described in Section~\ref{sec:phenoconstraints}. We show in Figure~\ref{fig:results-lfv} the constraints provided by LFV, being one of the most interesting. As with all the previous figures, the size of the dots represents the multiplicity of the models, whereas the colours indicate the constraint from the scalar contributions. As before, blue dots are unconstrained, while green, orange and red dots indicate dangerous contributions to LFVs if the relevant couplings are of orders $10^{-2} - 1$, $10^{-4} - 10^{-2}$ and $10^{-6} - 10^{-4}$, respectively. The gauge contribution is shown by levels of saturation and vertical lines. The low saturation to the left of the solid line shows excluded models, mid level of shading, between the solid and dashed lines, indicates prospective exclusion assuming one order of magnitude improvement, and full saturation means no exclusion. It can be easily noticed that the gauge contribution provides the strongest constraint in this case, effectively excluding models with $M_{LR} \lesssim 10^5$ GeV based on the current experimental limits. 

\begin{figure}[ht!]
 \centering
 \includegraphics[width=0.48\textwidth]{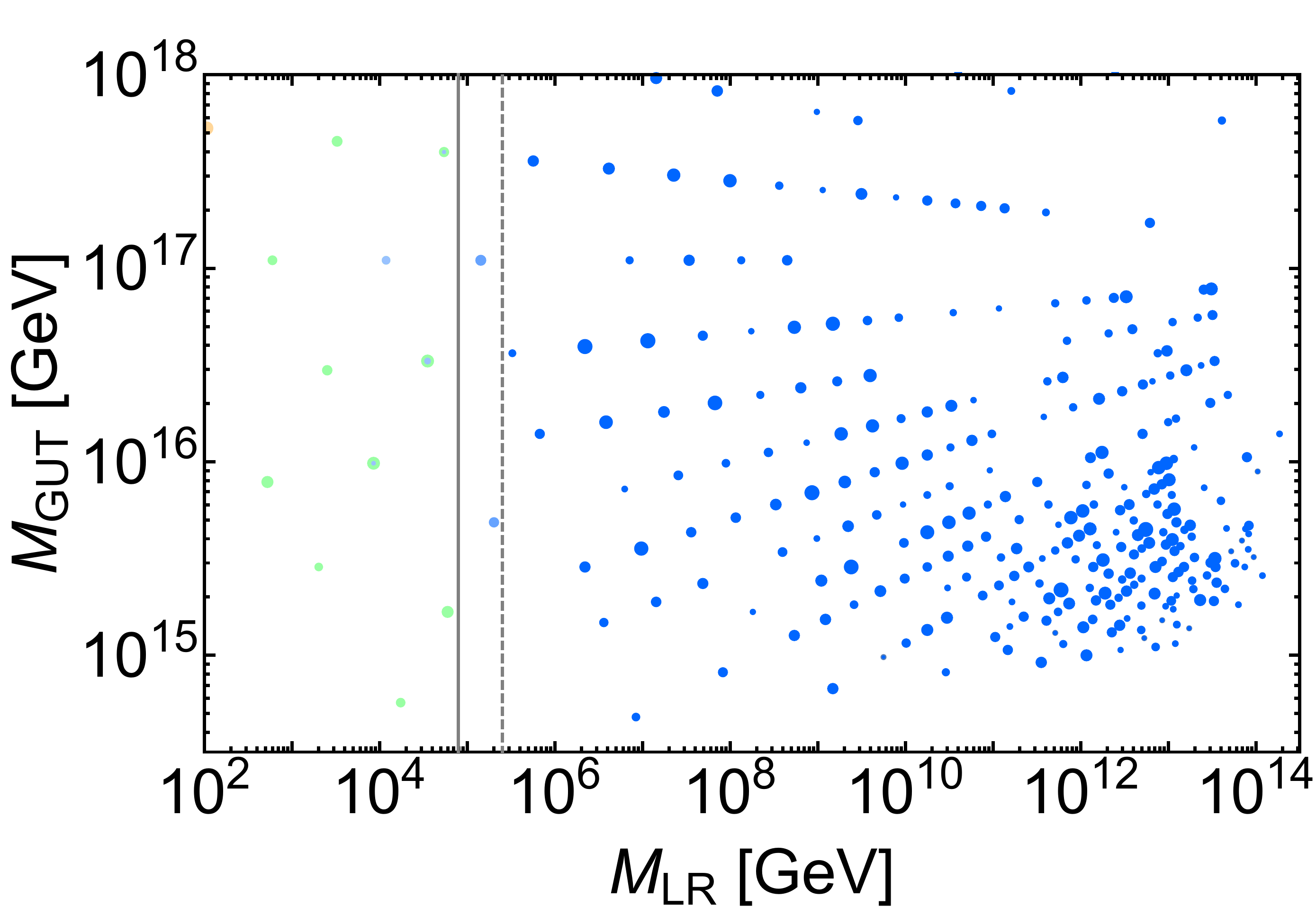}
 \includegraphics[width=0.48\textwidth]{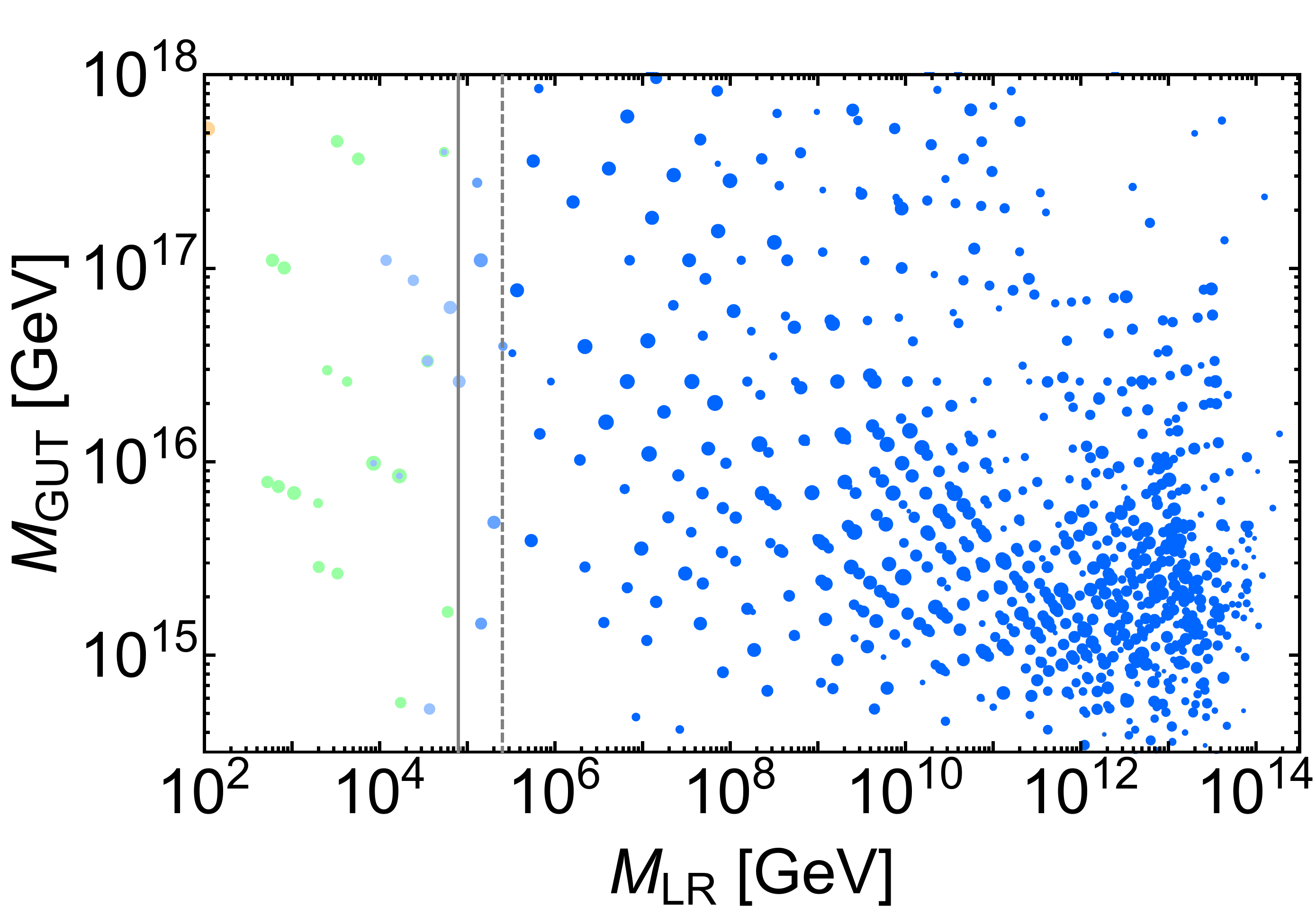}
 \caption{Effect of the LFV constraint on the solutions $(M_{GUT}, M_{LR})$ of the unified RG running for models with no exotic fields at the SM scale (left) and with only electroweak doublets and/or singlets at the SM scale (right). The colours represent the potential contribution of exotic scalar fields in the given model to LFVs. Shading and vertical lines represent the contribution from gauge interactions. The point sizes indicate the multiplicity of models.}
\label{fig:results-lfv}
\end{figure}

Lastly, in Figure \ref{fig:results-combined}, we show the combined effect of all observables described in Section \ref{sec:phenoconstraints}. The sizes, colours and shading of the dots, as well as horizontal an vertical lines,  have the same meaning as in all figures above. In addition, the thin dotted vertical lines mark out the mass scale of the right handed neutrino that satisfies the constraint in Eq. \ref{neutrinomasses} for a coupling, from right to left, of $ \lambda_\nu = \sqrt{\lambda_\Delta - y_\nu^2} \sim 10^0$, $10^{-2}$, $10^{-4}$ and $10^{-6}$. Obviously models with large $M_{LR}$ can predict the right neutrino masses with a reasonabe coupling of order $\lesssim 1$, however models with low LR scale may also be consistent with neutrino masses, if some level of fine tunning between the couplings $\lambda_\Delta$ and $y_\nu$ is assumed .If one takes the maximal exclusion allowed by these constraints, with all the assumptions that it takes, only a small region of the energy scale space is left, with $M_{GUT} \gtrsim 10^{16}$ GeV and $10^{14} \gtrsim M_{LR} \gtrsim 10^{5}$ GeV.

\begin{figure}[ht!]
 \centering
 \includegraphics[width=0.48\textwidth]{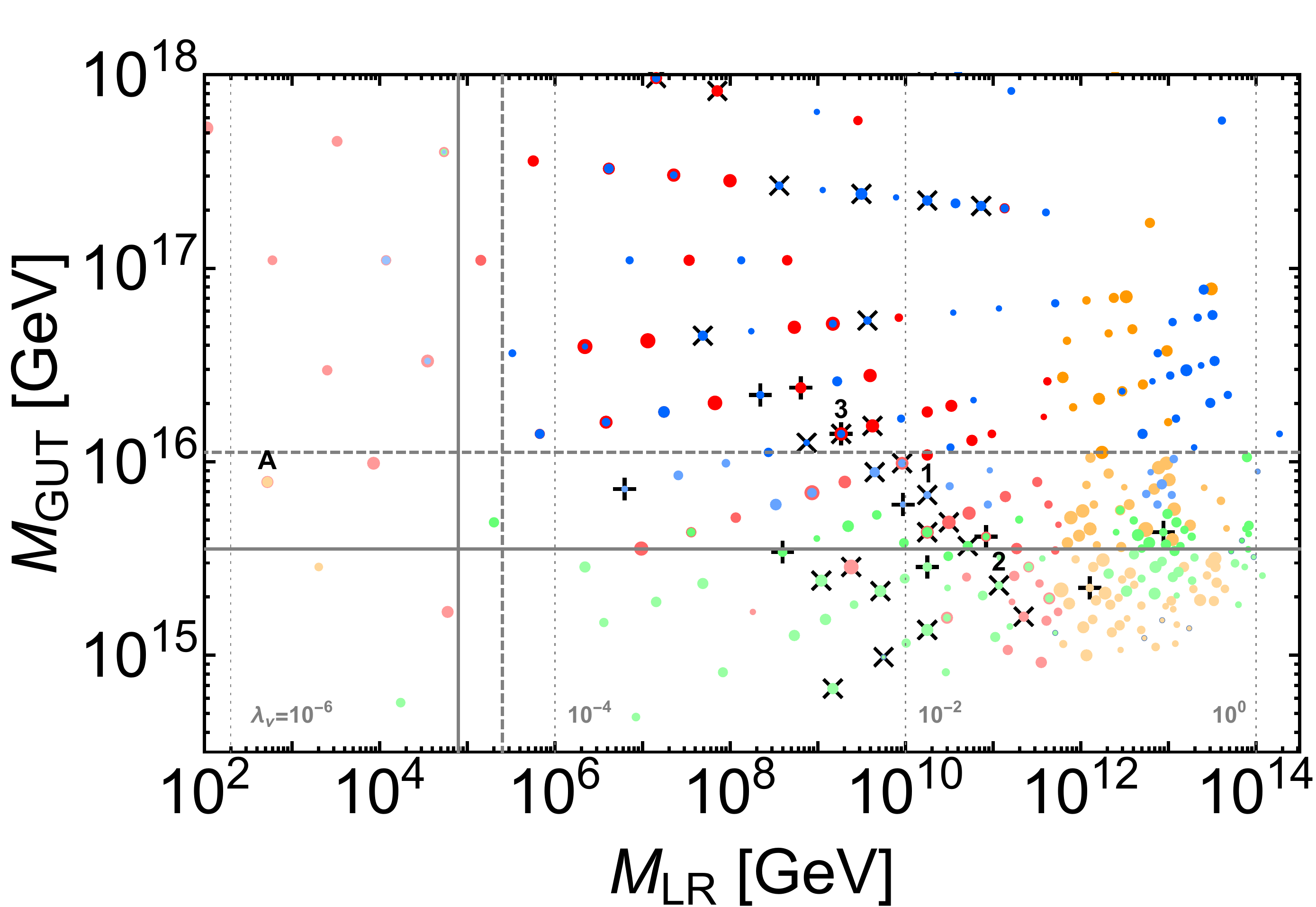}
 \includegraphics[width=0.48\textwidth]{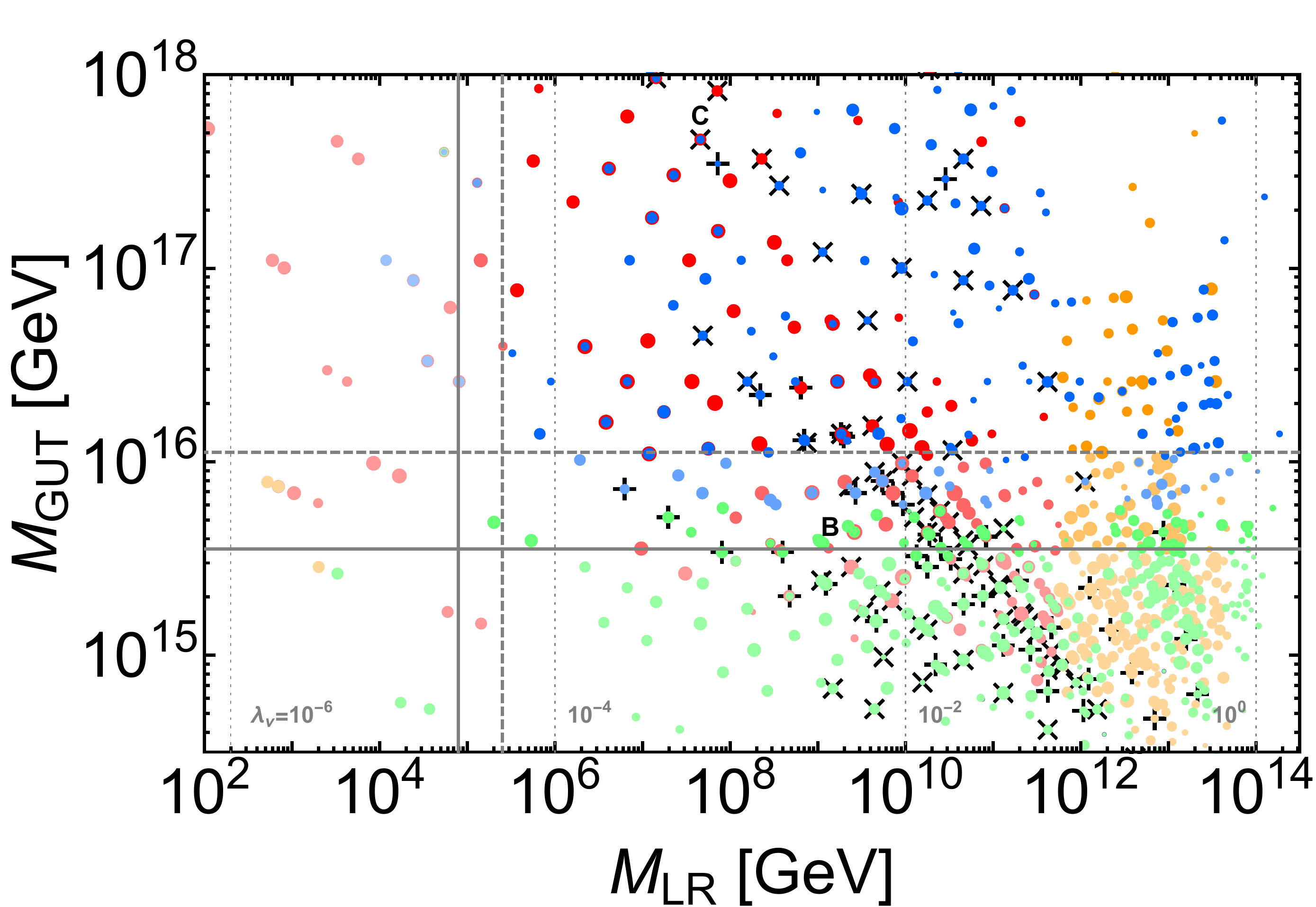}
 \caption{Combined effect of all observables on the solutions $(M_{GUT}, M_{LR})$ of the unified RG running for models with no exotic fields at the SM scale (left) and with only electroweak doublets and/or singlets at the SM scale (right). Size, colour, shading and lines have the same meaning as in Figures \ref{fig:results-protondecay} and \ref{fig:results-lfv}. As above, the crosses identify models with manifest left-right symmetry ($\times$) and models that satisfy the CKM constraint ($+$).}
\label{fig:results-combined}
\end{figure}

Among the many models obtained through the procedure described above, we will showcase a few example scenarios to illustrate our approach. The first of these (A), chosen to have just the SM particle content (left-hand plots in the figures above), has the following scalar representations at $M_{LR}$ and $M_{SM}$ scales
\begin{align}
 \notag \mathcal{R}_{LR}^A &\supset  \{\mathbf{1},\mathbf{2},\mathbf{2},~0\}\,, \{\mathbf{3},\mathbf{2},\mathbf{2},~\tfrac{1}{2}\}\,, \{\mathbf{\bar{3}},\mathbf{2},\mathbf{2},~1\} \,, \{\mathbf{\bar{3}},\mathbf{1},\mathbf{3},\tminus\tfrac{1}{2}\}\,, \{\mathbf{1},\mathbf{1},  \mathbf{3},\tminus\tfrac{3}{2}\}, \\
 \mathcal{R}_{SM}^A &\supset  \{\mathbf{1},\mathbf{2},\tminus\tfrac{1}{2}\}.
 \label{modelAreps}
\end{align}
The scalar field content at the LR scale includes an $SU(2)$ bidoublet, which contains the SM Higgs doublet, a $SU(2)_R$ triplet (responsible for LR symmetry breaking) and additional scalar representations. With this representation content given, one can calculate the running of the gauge couplings. The RGE running for this model is shown on the left hand plot of Figure \ref{fig:results-RGEs}. As one can see, the LR scale is $M_{LR} \sim 10^{2-3}$ GeV and GUT scale is $M_{GUT} \sim 10^{16}$ GeV. The low value of $M_{LR}$ lies within the experimental sensitivity of observables such as LFV (as can be noticed in Figure \ref{fig:results-combined}) which makes this scenario particularly interesting, since it can be probed by current and future experiments and possibly constrain the values of the LFV couplings.

The second scenario that we study (B) corresponds to the case of high LR scale, with representation content
\begin{align}
 \notag \mathcal{R}_{LR}^B &\supset  \{\mathbf{1},\mathbf{2},\mathbf{2},~0\}\,, \{\mathbf{3},\mathbf{2},\mathbf{2},~\tfrac{1}{2}\}\,, \{\mathbf{1},\mathbf{2},\mathbf{2},~0\} \,, \{\mathbf{1},\mathbf{1},\mathbf{3},\tminus\tfrac{3}{2}\}\,, \{\mathbf{1},\mathbf{1},  \mathbf{3},~\tfrac{3}{2}\}, \\
 \mathcal{R}_{SM}^B &\supset  \{\mathbf{1},\mathbf{2},~\tfrac{1}{2}\} \,, \{\mathbf{1},\mathbf{2},\tminus\tfrac{1}{2}\}\,, \{\mathbf{1},\mathbf{2},~\tfrac{1}{2}\} \,, \{\mathbf{1},\mathbf{2},\tminus\tfrac{1}{2}\}.
 \label{modelBreps}
\end{align}
This model has several exotics at the SM scale, all of them $SU(2)_L$ doublets, besides the SM Higgs. In this type of two Higgs doublet models (2HDM) one expects the lightest of the mixed states to correspond with the measured SM Higgs field, whereas the rest of the states, typically much heavier, might be found by future searches at colliders. Additionally, the large value for the LR scale, $M_{LR} \sim 10^{9}$ GeV, is typically beyond current experimental sensitivity, but it could perhaps be achieved with future experiments.

\begin{figure}[ht!]
 \centering
 \includegraphics[width=0.48\textwidth]{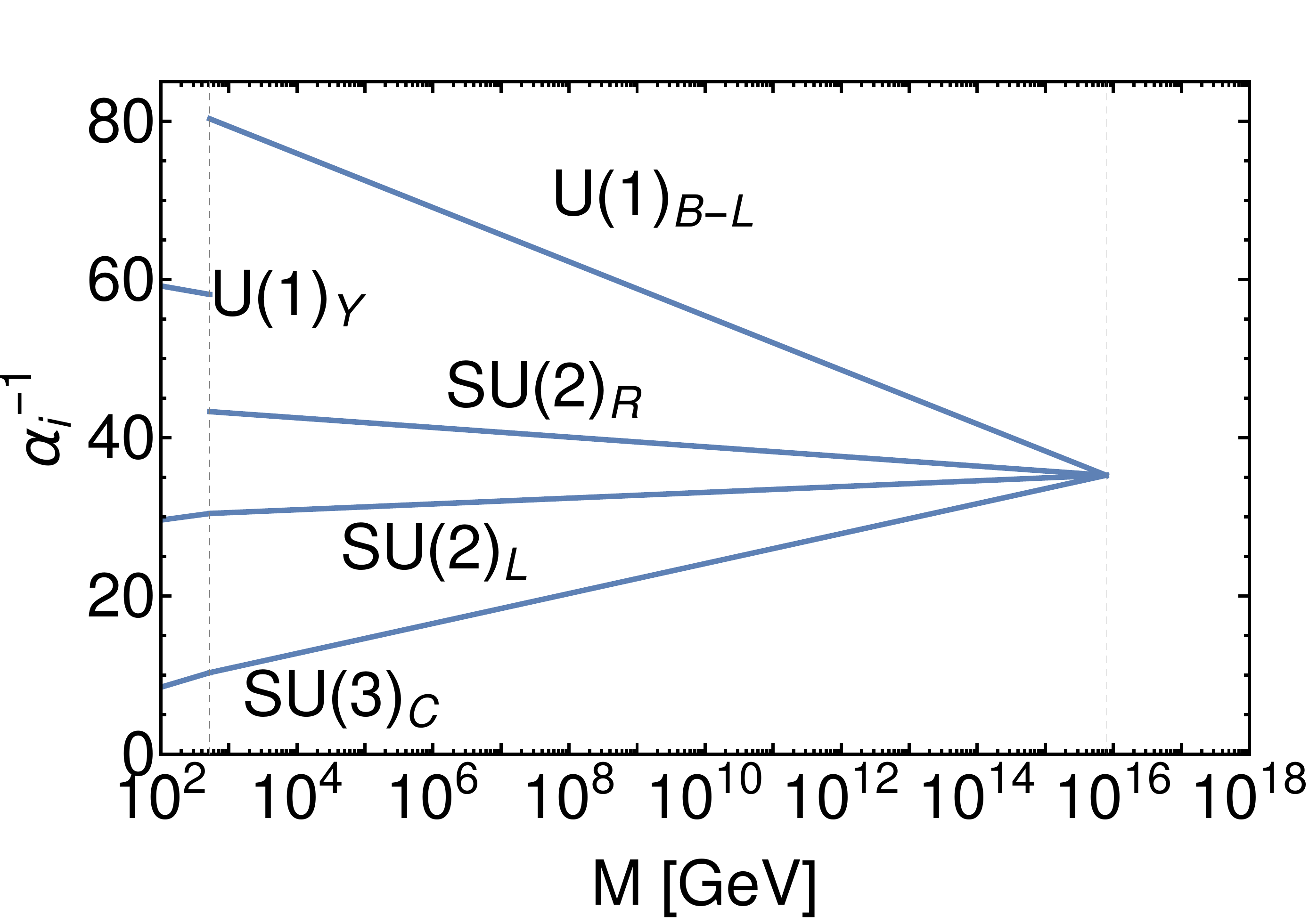}
 \includegraphics[width=0.48\textwidth]{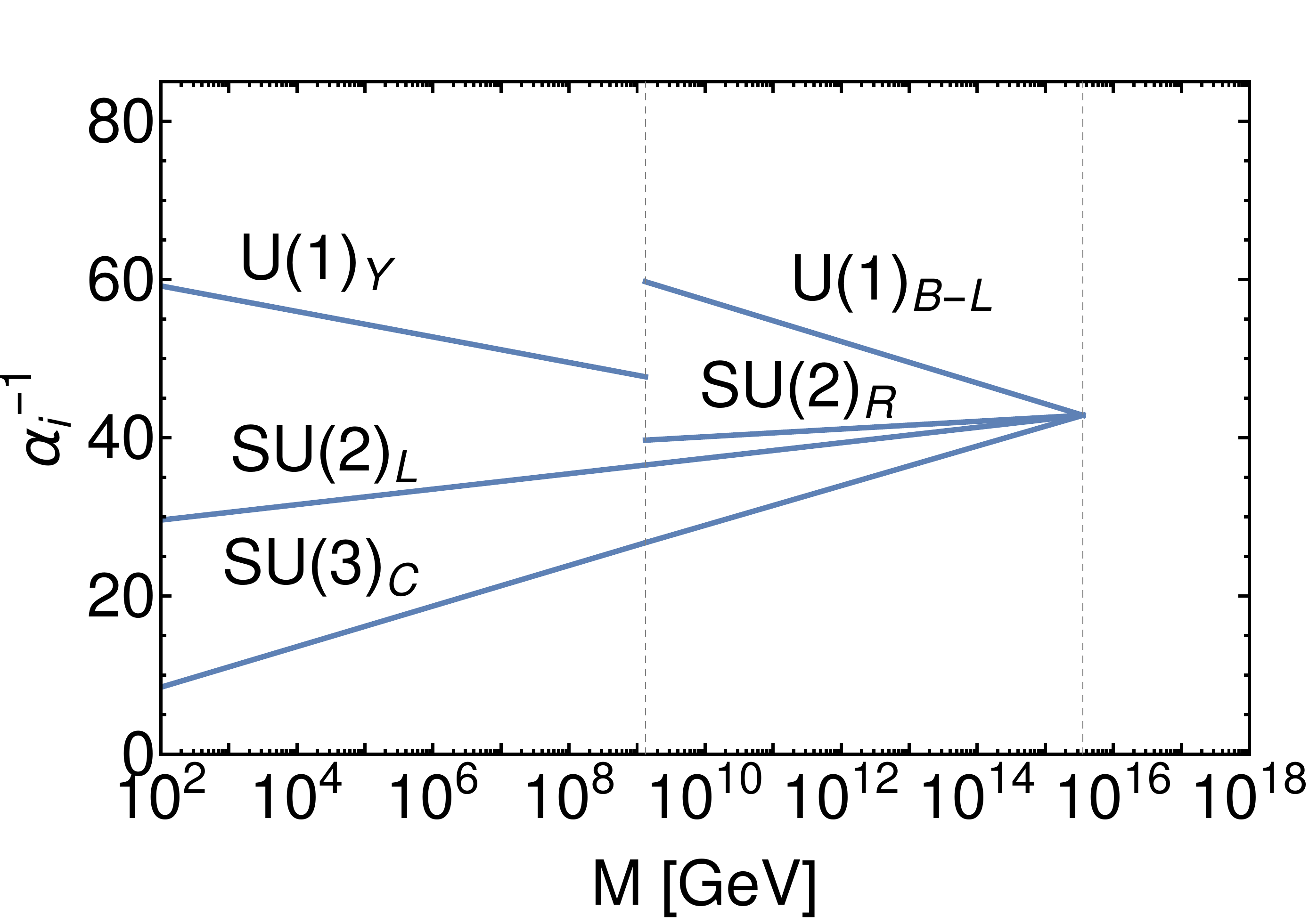}\\
 \includegraphics[width=0.48\textwidth]{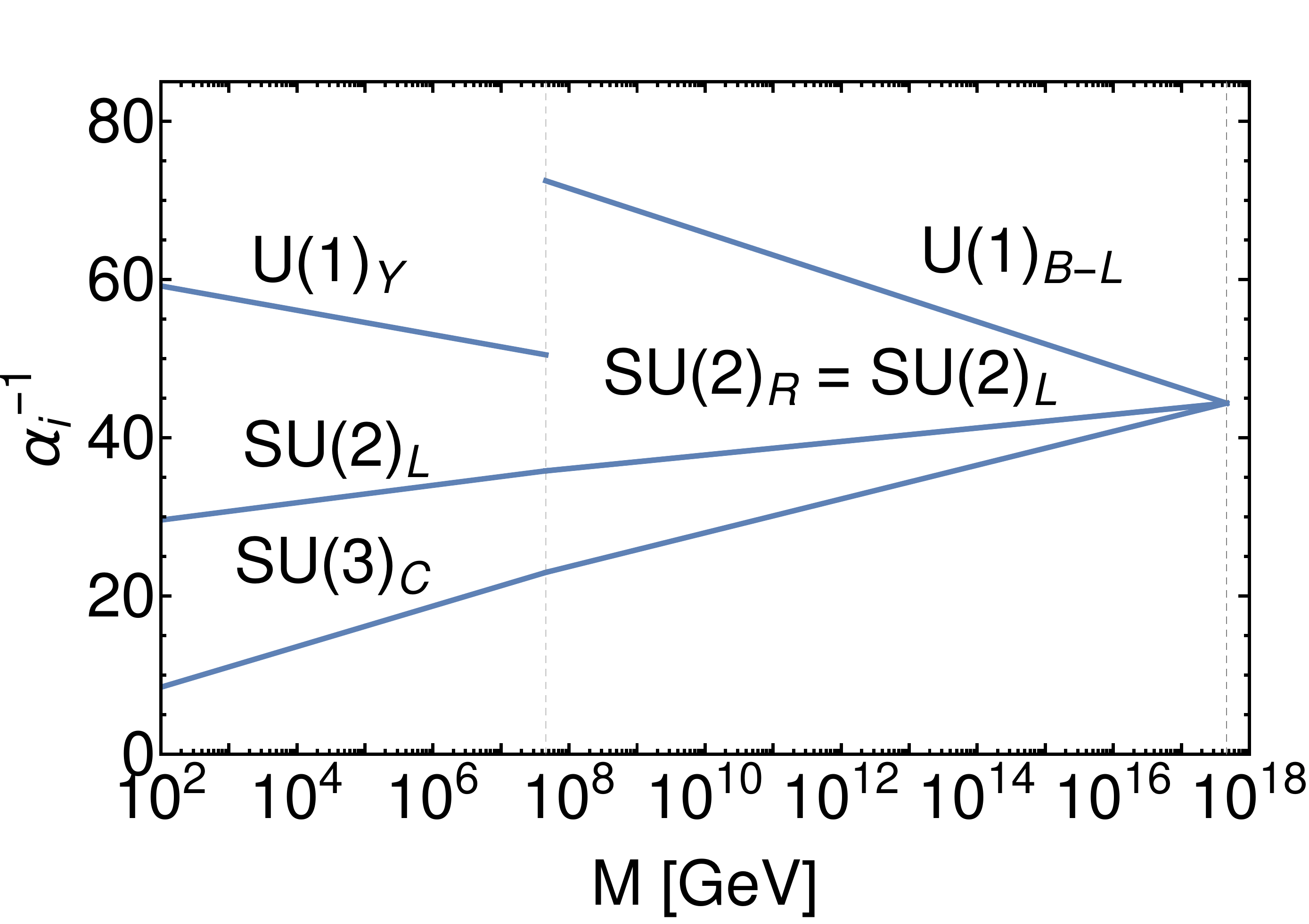}
 \caption{RGE running in the three example scenarios described in the text: low LR scale and only SM scalar particle content (A), high LR scale and extended SM particle content (B) and manifest LR scenario (C).}
\label{fig:results-RGEs}
\end{figure}

The last example model (C) will be manifestly left-right symmetric. Models A and B, with representation content in Eqs. \eqref{modelAreps} and \eqref{modelBreps}, respectively, are not manifestly LR symmetric, since both of them include $SU(2)_R$ triplets, but not the $SU(2)_L$ equivalents. In addition, in Figure \ref{fig:results-RGEs} one can notice the different slopes corresponding to the $SU(2)_L$ and $SU(2)_R$ gauge couplings arising due to the asymmetry of the particle content. An example representation content that is manifestly LR symmetric is
\begin{align}
 \notag \mathcal{R}_{LR}^C &\supset \{\mathbf{1},\mathbf{2},\mathbf{2},~0\}\,,\{\mathbf{8},\mathbf{1},\mathbf{1},~0\} \,, \{\mathbf{\bar{3}},\mathbf{1},\mathbf{1},~1\} \,,\{\mathbf{1},\mathbf{1}, \mathbf{3},\tminus\tfrac{3}{2}\} \, \{\mathbf{1},\mathbf{3},\mathbf{1},~\tfrac{3}{2}\}, \\
 \mathcal{R}_{SM}^C &\supset  \{\mathbf{1},\mathbf{2},\tminus\tfrac{1}{2}\}\,,\{\mathbf{1},\mathbf{2},~\tfrac{1}{2}\}.
 \label{modelCreps}
\end{align}
The RGE running of this model is shown in the bottom of Figure \ref{fig:results-RGEs}, where one can notice the joint running of the $SU(2)_L$ and $SU(2)_R$ couplings, due to the manifest LR symmetry. This model presents a very high GUT scale $M_{GUT} \sim 10^{18}$ GeV, close to the Planck scale, therefore there might be some gravitational corrections that could affect the unification of the gauge couplings, as can be seen in \cite{Robinson:2005fj}. 

Finally, we would like to make a quick comparison with other large scope GUT analyses, such as the one in \cite{Arbelaez:2013nga}. Due to the nature of our analysis and the restrictions on the number of representations we have imposed, we cannot reproduce most of the models suggested in the reference. We can however reproduce the simplest of those, given in Table I of \cite{Arbelaez:2013nga}. These models have only the SM Higgs at the SM scale and the following representations at the LR scale\footnote{We must point out that model 2 is not exactly the same as the one in \cite{Arbelaez:2013nga}, where it shows two copies of the bidoublet. Since that would exceed out limit of 5 on the number of representations, we have decided to show a similar model with just one copy of the bidoublet.}
\begin{align}
 \notag \mathcal{R}_{LR}^1 &\supset \{\mathbf{1},\mathbf{2}, \mathbf{2}, ~0\}\,, \{\mathbf{1},\mathbf{1}, \mathbf{3}, \tminus2\}\,, \{\mathbf{1},\mathbf{3}, \mathbf{1}, \tminus2\}, \\
 \notag \mathcal{R}_{LR}^2 &\supset \{\mathbf{1},\mathbf{2}, \mathbf{2}, ~0\}\,, \{\mathbf{1},\mathbf{1}, \mathbf{3}, ~0\}\,, \{\mathbf{1},\mathbf{3}, \mathbf{1}, ~0\}\,, \{\mathbf{1},\mathbf{1}, \mathbf{3}, \tminus2\}\,, \{\mathbf{1},\mathbf{3}, \mathbf{1}, \tminus2\}, \\
 \mathcal{R}_{LR}^3 &\supset \{\mathbf{1},\mathbf{2}, \mathbf{2}, ~0\}\,, \{\mathbf{1},\mathbf{2}, \mathbf{2}, ~0\}\,,\{\mathbf{1},\mathbf{1}, \mathbf{3}, ~0\}\,,\{\mathbf{1},\mathbf{1}, \mathbf{3}, \tminus2\}.
\end{align}
These models are shown in the scatter plots in Figure \ref{fig:results-combined}, labelled by their corresponding numbers, $1,2$ or $3$, respectively.

%---------------------------------------------------------
\section{Conclusions}
\label{sec:conclusions}
%---------------------------------------------------------

In this article we discuss a framework to automatically generate models of grand unification. We have described in detail the procedure used for realising symmetry breaking and gauge coupling unification, focusing on unified theories with the $SO(10)$ gauge group. The aim of this framework is the generation of a large landscape of models, differing in the position of the energy scales which correspond to the symmetry breaking steps, and the field content at said energy scales. A number of theoretical constraints have been imposed so as to satisfy conditions such as chirality and anomaly cancellation, among others. Although the constraints imposed are but a subset of those required for a realistic broken gauge model at high energies, we have included those that can be checked without a detailed Lagrangian description.

The renormalization group running of the gauge couplings was thoroughly described, highlighting the analytical approach to obtain the solution of the one-loop gauge RGEs. It includes the case of symmetry breaking of multiple Abelian groups with mixing of Abelian charges and couplings. Appropriately for the level of accuracy needed for our purposes, we only use one-loop RG running and we accordingly neglect threshold corrections.

We have applied this procedure to an $SO(10)$ model with an intermediate left-right symmetry $SU(3)_c \times SU(2)_L \times SU(2)_R \times U(1)$, which we consider as one of the most interesting breaking chains. For this particular scenario, we have calculated the approximate contribution of several low energy observables, such as neutrino masses, proton decay, neutron oscillations and rare lepton flavour violating decays. In the results we have shown how the landscape of models is distributed over the $M_{LR}-M_{GUT}$ scale plane, along with the constraints from current and future experimental measurements. 

GUT models in general and SO(10) GUTs in particular remain well-motivated scenarios of BSM physics, even without the discovery of supersymmetry close to the electroweak scale. The main goal of this work is to go beyond the (possibly unjustified) bias of minimality in GUT models and survey the richness of scenarios extended by additional exotic scalar fields, but also to provide an initial pathway to use experimental probes to constrain solutions. Clearly, allowing even a small number of exotic representations to survive at the LR scale or even the SM scale permits a very high number of solutions. Most of these occur for large LR scales $\approx 10^{10-12}$~GeV and fairly canonical GUT scales $\approx 10^{15-16}$~GeV. This is largely a consequence of the fact that the SM gauge couplings converge around such energies, and the inclusion of the intermediate LR scale and additional exotics admits exact unification in many cases. Within our approach, we do not find any successful scenarios with LR scales above $\approx 10^{14}$~GeV. Many scenarios contain coloured exotic leading to potentially rapid already proton decay searches. This is for example the case for many scenarios within the window $M_{LR} \approx 10^{12-13}$~GeV which are therefore ruled out unless the associated couplings are very small (see the yellow coloured points in Fig.~\ref{fig:results-combined}). There remain only a few viable solutions with LR scales low enough to be testable in laboratory experiments.

We would like to point out that this is but the first of a series of analyses on GUT models. The mechanism described can be applied to any unified gauge group with any breaking chain (satisfying the relevant conditions). Therefore, we expect to be able to perform an automatic analysis similar to this one also for other scenarios, such as the Pati-Salam model or an $SU(5)$-inspired model. Lastly, we acknowledge that some of the decisions made in order to simplify the generation of models are, at best, reasonable approximations. Further refinements and modifications can be implemented later, e.g. in deeper theoretical contexts. In this sense, the results shown may serve as an intermediate step for the analysis of a more comprehensive GUT theory.

%---------------------------------------------------------
\section*{Acknowledgements}
%---------------------------------------------------------
The work of FFD and TEG was supported partly by the London Centre for Terauniverse Studies (LCTS), using funding from the European Research Council via the Advanced Investigator Grant 267352. TEG was also partly funded by the Research Council of Norway under FRIPRO project number 230546/F20.

\appendix

%%%%%%%%%%%%%%%%%%%%%%%%%%%%%%%%%%%%%%%%%%%%%%%%%%%%%%%%%%

\bibliographystyle{h-physrev4}
\bibliography{so10_survey}

\end{document}